\documentclass[12pt]{article}
\usepackage{amsmath,amsfonts}
\usepackage{hyperref}
\usepackage{graphicx}
\usepackage{feynmp}
\usepackage{xcolor}
\usepackage{dsfont}
\usepackage{cite}
\usepackage{amsthm}
\usepackage{amssymb}
\usepackage[matrix,arrow]{xy}
\usepackage{epsfig}
\usepackage{dsfont}

\makeatletter\@addtoreset{equation}{section}\makeatother

\def\bR {\mathbb{R}}

\newcommand{\fs}{\!\!\!/}
\newcommand{\D}{\nabla\!\fs}
\newcommand{\deq}{\overset{\mathrm{mod}\,\delta}{=}}
\newcommand{\rf}[1]{(\ref{#1})}
\newcommand{\beq}{\begin{equation}}
\newcommand{\eeq}{\end{equation}}
\newcommand{\bea}{\begin{eqnarray}}
\newcommand{\eea}{\end{eqnarray}}

\newcommand{\vev}[1]{{\left< {#1} \right>}}

\newcommand{\CO}{{\cal O}}
\newcommand{\CN}{{\cal N}}

\newcommand{\onov}[1]{\frac{1}{#1}}
\newcommand{\eq}[2][ ]{\begin{equation}\label{#1}{\begin{split}#2\end{split}}\end{equation}}
\newcommand{\eql}[2]{\begin{equation}\label{#1}{\begin{split}#2\end{split}}\end{equation}}
\newcommand{\mat}[1]{\left(\begin{matrix} #1 \end{matrix}\right)}

\newcommand{\Tr}{{\rm Tr\,}}

\newcommand{\cC}{{\mathcal C}}
\newcommand{\cE}{{\mathcal E}}

\newcommand{\cF}{{\mathcal F}}

\newcommand{\cL}{{\mathcal L}}
\newcommand{\cM}{{\mathcal M}}
\newcommand{\cN}{{\mathcal N}}

\newcommand{\cQ}{{\mathcal Q}}

\newcommand{\cO}{{\mathcal O}}

\newcommand{\dbar}{\partial_{\bar{\tau}}}
\newcommand{\dtau}{\partial_\tau}

\renewcommand{\title}[1]{\vbox{\center\LARGE{#1}}\vspace{5mm}}
\renewcommand{\author}[1]{\vbox{\center\large#1}\vspace{5mm}}

\begin{document}
\bibliographystyle{utphys}
\begin{fmffile}{graphs}

\begin{titlepage}
\begin{center}
%\vspace{5mm}
\hfill {\tt PUPT-2500}\\
\hfill {\tt }\\
\vspace{8mm}

\title{Correlation Functions of Coulomb Branch Operators}
\vspace{4mm}

\end{center}

\author{Efrat Gerchkovitz,$^1$ Jaume Gomis,$^2$ Nafiz Ishtiaque,$^{2,3}$ Avner Karasik,$^1$ \\[+0.5ex] Zohar Komargodski,$^1$ and Silviu S.~Pufu$^4$} 
\vspace{3mm}
\begin{center}
\noindent
$^1$   Weizmann Institute of Science, Rehovot 76100, Israel \\[1ex]
$^2$   Perimeter Institute for Theoretical Physics, Waterloo, ON N2L 2Y5, Canada \\ [1ex]
$^3$   Department of Physics, University of Waterloo, Waterloo, ON N2L 3G1, Canada\\[1ex]
$^4$  Joseph Henry Laboratories, Princeton University, Princeton, NJ 08544, USA 
\end{center}
\vspace{5mm}
\abstract{\normalsize{We consider the correlation functions of Coulomb branch operators in four-dimensional $\mathcal{N}=2$ Superconformal Field Theories (SCFTs) involving exactly one anti-chiral operator.  These extremal correlators are the ``minimal" non-holomorphic local observables in the theory. We show that they can be expressed in terms of certain determinants of derivatives of the four-sphere partition function of an appropriate deformation of the SCFT\@.  This relation between the extremal correlators and the deformed four-sphere partition function is non-trivial due to the presence of conformal anomalies, which lead to operator mixing on the sphere.  Evaluating the deformed four-sphere partition function using supersymmetric localization, we compute the extremal correlators explicitly in many interesting examples.  Additionally, the representation of the extremal correlators mentioned above leads to  a system of integrable differential equations. We compare our exact results with previous perturbative computations and with the four-dimensional $tt^*$ equations. We also use our results to study some of the asymptotic properties of the perturbative series expansions we obtain in $\mathcal{N}=2$ SQCD.}

\noindent
}
\vfill

\end{titlepage}

\tableofcontents

%%%%%%%%%%%%%%%%%%%%%%%%%%%%%%
\vfill\eject

\vfill
%%%%%%%%%%%%%%%%%%%%%%%%%%%%%%%%%%%%%%%%%%%%%%%%%%
\section{Introduction and Conclusions} 
\label{sec:intro}
 
The correlation functions of local operators are amongst the most well-studied observables in Quantum Field Theory (QFT)\@. 
In Conformal Field Theories (CFTs),  the two- and three-point functions of  the  local operators in the theory  completely determine all the  $n$-point functions.    
\smallskip

In this paper we  find a  formula that computes  exactly the correlation functions\footnote{Chiral primary operators sit in short representations of the four dimensional $\cN=2$ superconformal algebra. See section \ref{sec:cpodet}.}
\begin{equation}
\label{extremalo}
\vev{\cO_{I_1}(x_1)\cO_{I_2}(x_2)...\cO_{I_n}(x_n){\overline \cO}_{\bar J}(y)}_{\bR^4} 
\end{equation}
of any number of chiral primary operators $\cO_{I_i}$ and one anti-chiral primary operator $\overline\cO_{\bar J}$ in four-dimensional $\cN=2$ superconformal field theories (SCFTs). Such correlation functions are henceforth referred to as extremal correlators. We determine  these correlators as   functions of the exactly marginal couplings of the SCFT, which span the so-called conformal manifold of the SCFT.\footnote{See section \ref{sec:confman} for more details.} Our results apply to any   SCFT with exactly marginal couplings that admit a Lagrangian description somewhere on the conformal manifold.\footnote{The structure we find  also applies to SCFTs that are inherently non-Lagrangian.} Near a weakly coupled point on the conformal manifold  we find that   the  correlators \rf{extremalo} are given by an infinite series of perturbative corrections dressed by an infinite sequence of nonperturbative instanton corrections.   Special cases of~\eqref{extremalo} are  the 
two- and three-point functions, which we refer to as the ``chiral ring data'' of the SCFT\@.   As we will review below, once the chiral ring data is known, all the extremal correlators can be reconstructed.

 \smallskip
 
Roughly speaking, our strategy is to  express   the flat space ($\bR^4$) correlators given in \rf{extremalo} in terms of the four-sphere ($S^4$) partition function of a suitable {\it deformation} of the SCFT\footnote{Our essential ideas and techniques can be also applied to $(2,2)$ theories in $d=2$. However, we do not pursue this direction here and concentrate on   $\cN=2$ theories in $d=4$. In fact, technically the case of $d=2$ is   simpler since no new instanton contributions need to be computed in the Coulomb branch representation~\cite{Benini:2012ui,Doroud:2012xw}.} 
\beq
\label{slogan}
Z_{\text{deformed}}[S^4]\Longrightarrow \vev{\cO_{I_1}(x_1)\cO_{I_2}(x_2)\ldots \cO_{I_n}(x_n)\overline\cO_{\bar J}(y)}_{\bR^4} \,.
\eeq
This partition function can be in turn evaluated exactly by supersymmetric localization, expanding upon Pestun's computation of the undeformed $\cN=2$~partition function~\cite{Pestun:2007rz}.  As was the case with the undeformed sphere partition function studied by Pestun, $Z_{\text{deformed}}[S^4]$ is also expressed as integral over the norm of the deformed partition function in the $\Omega$-background~\cite{Nekrasov:2002qd} (see section \ref{sec:defpart}).
\smallskip

An important subtlety in the relation \eqref{slogan} between $Z_{\text{deformed}}[S^4]$ and the extremal correlators on $\bR^4$ is due to conformal anomalies, which cause operator mixing on $S^4$.  Diagonalizing the operator mixing matrix on $S^4$   \`a la Gram-Schmidt leads to a representation of the  extremal  correlators  on $\bR^4$ in terms of determinants of derivatives of the deformed sphere partition function   $Z_{\text{deformed}}[S^4]$.  This induces the action of a system of integrable differential equations on the extremal correlators of $\cN=2$ SCFTs. 

\smallskip

As an illustrative example, we can consider $SU(2)$ SQCD with $4$ fundamental hypermultiplets, which contains precisely one chiral primary operator of dimension $2n$ for every integer $n\geq 1$.  This case is special in that one does not have to consider any deformations of the $S^4$ partition function in order to calculate extremal correlators.  The two-point functions of the dimension $2n$ chiral primary operators $\cO_n$ can be  expressed succinctly as the ratio of determinants
\beq
\vev{\cO_n(0)\overline \cO_m(\infty)}_{\bR^4}= \frac{16^{n} \delta_{nm}}{Z[S^4]}{\det_{(k,l)=0,\ldots,n} \left(\partial_\tau^k \partial_{\overline\tau}^l Z[S^4]\right)\over\det_{(k,l)=0,\ldots,n-1}  \left(\partial_\tau^k \partial_{\overline\tau}^l Z[S^4]\right)}\,,
\label{generaldet}
\eeq
where $\tau$ is the complexified coupling constant of the theory. This formula neatly encodes all the two-point functions of chiral primary operators in terms of the sphere partition function, which can be computed exactly by supersymmetric localization.
  \smallskip

The partition function $Z_{\text{deformed}}[S^4]$, and therefore the extremal correlators, can be explicitly calculated to all orders in perturbation theory. 
The   instanton corrections  to  $Z_{\text{deformed}}[S^4]$   can be computed in some theories using results already available in the literature, while for other theories it requires first writing down the instanton partition function of the deformed SCFT in the $\Omega$-background, which is an interesting open problem (see section \ref{sec:defpart}).

\smallskip
Our identification, summarized by the schematic equation \rf{slogan}, provides a broad extension of the formula derived in~\cite{Gerchkovitz:2014gta,Gomis:2014woa,Gomis:2015yaa} relating the undeformed $S^4$ partition function of the SCFT to the K\"ahler potential $K$ on the conformal manifold\footnote{Extending the earlier result in two-dimensional $\cN=(2,2)$ SCFTs~\cite{Jockers:2012dk,Gomis:2012wy,Gerchkovitz:2014gta,Gomis:2015yaa}.}   
\begin{equation}\label{PartitionFun}Z[S^4]=r^{-4a}e^{{1\over 12}K(\tau^i,\bar\tau^{\bar i})}~,\end{equation} 
where $\tau^i, \bar\tau^{\bar i}$ are the exactly marginal couplings of the SCFT, $a$ is the Euler conformal anomaly and $r$ the radius of $S^4$. The two-point functions of the dimension-two chiral primary operators, denoted by $\cO_i$, are   determined in terms of the $S^4$ partition function of the SCFT through
\beq
 \vev{\cO_i(0)\overline\cO_{\bar i}(\infty)}_{\bR^4} = 16  {\partial\over \partial \tau^i} {\partial\over \partial \bar \tau^{\bar i}}\ln Z[S^4]\,.
\label{flatcorrelaxxx}
\eeq
Our results extend this  formula  to arbitrary chiral primary operators $\cO_I$. See, for example, equation \rf{generaldet}.

\smallskip
As mentioned above, the chiral ring data  obtained from the deformed partition function $Z_{\text{deformed}}[S^4]$ obeys a system of differential equations     with respect to the exactly marginal couplings $\tau^i, \bar\tau^{\bar i}$. For  SCFTs with one exactly marginal coupling  and a one-dimensional Coulomb branch, namely for $\cN=2$ $SU(2)$ SQCD  with four fundamental hypermultiplets and $\cN=4$ $SU(2)$ super-Yang-Mills, we show that the equations obeyed by  the chiral ring data obtained from $Z_{\text{deformed}}[S^4]$ are those of a semi-infinite Toda chain, which are integrable.
 
\smallskip
The fact that the chiral ring data of these theories obeys the semi-infinite Toda chain system was exhibited in~\cite{Baggio:2014ioa,Baggio:2014sna,Baggio:2015vxa} starting from the 
the $tt^*$ equations of the four-dimensional SCFT~\cite{Papadodimas:2009eu}.
In Appendix \ref{ap:int} we show that the $tt^*$ equations of any four-dimensional $\cN=2$ SCFT   
are   integrable and   governed by a Hitchin system, in parallel with  the $tt^*$ equations of two-dimensional $(2,2)$ QFTs~\cite{Cecotti:1991me}.
In Appendix~\ref{couptt} we show that the chiral ring data of $SU(N)$ SQCD with $2N$ fundamental hypermultiplets computed through our correspondence~\eqref{slogan} indeed  obeys the corresponding $tt^*$ equations. For the special case of $\cN=4$   super-Yang-Mills with an  arbitrary gauge group $G\neq SU(2)$, the chiral ring can be organized in terms of decoupled semi-infinite Toda chains. However, this is not the case in $SU(N)$ SQCD with $2N$ fundamental hypermultiplets.

 \smallskip
 
The $tt^*$ equations  themselves are not sufficient to determine the chiral ring data of the SCFT since these equations have several solutions. Rather,  the chiral ring data is  found through the partition function of the deformed SCFT on $S^4$ via~\rf{slogan}. One can view~\rf{slogan} as a particular solution to the $tt^*$ equations. 
This allows us to obtain new results in four-dimensional $\cN=2$ SCFTs.

\smallskip
 The  computation of the correlation function of local operators~\rf{extremalo} in a four-dimensional QFT contributes to the recent progress in the   exact determination of certain
observables in supersymmetric QFTs. Particularly striking are those observables that depend non-holomorphically on the coupling constants  of the theory. These include the computation of   Wilson loops ~\cite{Pestun:2007rz}, {'t~Hooft} loops~\cite{Gomis:2011pf}, domain walls~\cite{Drukker:2010jp,Hosomichi:2010vh} and cusp anomalous dimensions at small angles~\cite{Fiol:2015spa} in four dimensional $\cN=2$ QFTs.  
For some previous work on  the  partition function of SCFTs on spheres consult~\cite{Pestun:2007rz,Kapustin:2009kz,Jafferis:2010un,Hama:2010av,Benini:2012ui,Doroud:2012xw,Doroud:2013pka}.

\smallskip 
The extremal correlators~\rf{extremalo}   should  transform    under the action of dualities. Indeed, a  chiral ring operator  is expected to transform as a modular form  under S-duality, with the modular weight determined by the dimension of the operator (c.f. \cite{Gukov:2006jk,Gomis:2009xg}). It would be interesting to study in detail the action of duality on these correlation functions. The exact computation of the extremal correlators in this paper can be  generalized by   adding supersymmetric circular Wilson loops,  't Hooft loops and/or  domain walls supported on $S^3$ in $\bR^4$, to yield, for example, the correlators\footnote{One could   also compute correlators in the presence of a surface operator by figuring out the interplay between vortices and instantons with the higher  dimensional chiral primary operators.}
\beq
\vev{\cO_I(0) D\, \overline\cO_{\bar J}(\infty)}_{\bR^4}\,,
\eeq
where $D$ denotes a judiciously chosen supersymmetric spherical defect operator in the SCFT. 

\smallskip

The results of our work are complementary to those coming from the superconformal bootstrap of 4d ${\cal N} = 2$ theories \cite{Beem:2014zpa, Liendo:2015ofa, Lemos:2015orc, Lemos:2015awa}.  In particular, \cite{Beem:2014zpa, Lemos:2015awa} considered four-point correlation functions of two chiral and two anti-chiral operators and obtained bounds on various OPE coefficients including some that can be computed from extremal correlators.  Their results pertaining to dimension-2 chiral primaries can be interpreted as bounds on the curvature of the conformal manifold.  It would be interesting to extract bounds on the curvature of the bundles of higher-dimension conformal operators in a similar way.

\smallskip 
The plan of the rest of the paper is as follows. In the remaining of the present section  we provide some relevant preparatory material: a brief discussion of conformal manifolds in CFTs, a review of the chiral ring of four-dimensional $\cN=2$ SCFTs, and a discussion of some subtleties that arise in defining CFTs on $S^4$.  In section \ref{sec:ring} 
 we show that  the chiral ring data of a SCFT can be extracted from the partition function of a deformation of the SCFT on $S^4$, and we provide an algorithm to determine the Hermitian metric on the chiral ring.  In section~3  we study   in detail $SU(N)$ SQCD and $\mathcal{N}=4$ super-Yang-Mills  and discuss the relation with the four-dimensional $tt^*$ equations.  We also consider some of the asymptotic properties of the perturbative expansion in $SU(N)$ SQCD. Many technical results are collected in five appendices.

\subsection{Conformal Manifolds}
 \label{sec:confman}
Let us review very briefly the notion of a conformal manifold. Given a CFT in $d$ dimensions, we suppose that there exists a (Hermitian) scalar marginal operator, $O$. If we deform the theory by $\delta S= \lambda \int d^dx\, O$ with some coefficient $\lambda$, then in general there would be a nontrivial beta function for $\lambda$ computable in conformal perturbation theory
\begin{equation}\label{BETA}{d\lambda\over d\ln\mu}= \beta_1\lambda^2+\beta_2\lambda^3+\cdots~.\end{equation}
However, under some circumstances, all the coefficients vanish $\beta_{a}=0$. We then say that $O$ is an exactly marginal operator; adding it to the action does not break the conformal symmetry. The coupling $\lambda$ in this case defines a line of CFTs along which the critical exponents can vary continuously. More generally, imagine that there is a set of such exactly marginal operators $O_i$. We can define the Zamolodchikov 
metric~\cite{Zamolodchikov:1986gt} in the space of theories,  that is, in the conformal manifold, via
\begin{equation}\label{Zamometric}\langle O_i(x)O_j(0)\rangle_{\{\lambda^i\}}={g_{ij} (\lambda^i)  \over x^{2d}} ~,\end{equation}
where we evaluate the two-point function in the CFT with couplings $\lambda^i$.
While the metric itself is as usual ambiguous (by choosing appropriate contact terms for our operators, we can choose the metric and the
Christoffel symbols to be trivial at any given point~\cite{Kutasov:1988xb}), there are various invariants such as the Ricci scalar that can be constructed out of it, and which are interesting observables of the CFT.

\smallskip

 The vanishing of all the coefficients $\beta_a=0$ in~\eqref{BETA} is common in $c=1$ models in $d=2$  but otherwise requires new symmetries in addition to the conformal symmetry~\cite{Cardy:1987vr}. One such extra symmetry is current algebra symmetry, in which case the spectrum of exactly marginal operators can be determined \cite{Forste:2003km}.
Another additional symmetry is supersymmetry. Indeed, exactly marginal operators are common in supersymmetric theories in $2\leq d\leq 4$. Let us consider first $\mathcal{N}=1$ theories in $d=4$. In these theories the conformal manifold is a K\"ahler manifold with local complex coordinates $\tau^{i}$, $\bar \tau^{\bar i}$ associated to the descendants of $\mathcal {N}=1$ chiral primaries and anti-chiral primaries of dimension 3. Not every marginal operator is necessarily exactly marginal, but there are nevertheless many examples with exactly marginal operators~\cite{Leigh:1995ep, Green:2010da}. $\mathcal {N}=2$ theories, being a special case of $\mathcal{N}=1$ theories, also admit a K\"ahler conformal manifold and the complex coordinates $\tau^i$,$\bar\tau^{\bar i}$ correspond to descendants of $\mathcal{N}=2$ chiral primaries of dimension 2 (see section \ref{sec:cpodet}).\footnote{This is for exactly marginal operators that  preserve $\cN=2$  supersymmetry.}  In an $\cN=2$ theory every marginal operator is necessarily exactly marginal.\footnote{Here is an argument along the lines of~\cite{Green:2010da}. There is a scheme in which the superpotential is not renormalized. Then if the beta function is nonzero it has to be reflected by a $D$-term in the action $\int d^4x d^8\theta \ \mathcal{U}$ with $\mathcal{U}$ some real primary operator. But since the $\tau^i$ are classically dimensionless, $\Delta(\mathcal{U})=0$ in the original fixed point. Therefore, $\mathcal{U}$ has to be the unit operator and the deformation $\int d^4x d^8\theta \ \mathcal{U}$   is therefore trivial. This proves that $\beta_a=0$.}  
One can further argue that in $\mathcal{N}=2$ theories the K\"ahler class is trivial, in other words, there are no two-cycles in the conformal manifold through which the K\"ahler two-form has flux. This global restriction implies, for example, that the $\mathcal{N}=2$ conformal manifold cannot be compact~\cite{Gomis:2015yaa}.
 
 \smallskip

 In four-dimensional $\CN=2$  SCFTs the K\"ahler potential (and hence the Zamolodchikov metric)  on the conformal manifold    can be determined exactly from the partition function of the SCFT on $S^4$ via \rf{PartitionFun}. 

 \smallskip

In some theories, different points in the conformal manifold may be mapped into each other by a duality transformation, possibly relating the theory in a regime where perturbation theory is valid to a  strongly coupled regime. This picture can give rise to an intricate pattern of dualities, where the conformal manifold can   acquire an elegant geometrical and mathematical interpretation, as in \cite{Gaiotto:2009we}. 

\smallskip
The extremal correlators~\rf{extremalo} provide  novel  QFT data  that transforms naturally under dualities.  It would be interesting to study in detail the action of strong-weak coupling dualities on these extremal correlation functions.

\subsection{The Chiral Ring of $\mathcal{N}=2$ SCFTs}
\label{sec:cpodet}

Local operators in $\bR^4$ or equivalently states on the cylinder  in an $\cN=2$  SCFT fit into unitary highest weight representations of the     
 superconformal algebra $su(2,2|2)$. The algebra $su(2,2|2)$ contains the following generators (in Euclidean signature):
 
 \begin{itemize}
\item The   conformal algebra $so(5,1)$   
\item The Poincar\'e supercharges $Q_\alpha^a, \overline Q_{\dot\alpha}^{\,a}$  and the  conformal supercharges $S_\alpha^a, \overline S_{\dot\alpha}^{\,a}$~~ ($a=1,2$)  
\item  The $su(2)_R\times u(1)_R$ R-symmetry  
 \end{itemize}
The (anti)-commutation relations can be found, for example, in~\cite{Freedman:2012zz}.
  \smallskip

A highest weight representation  is labeled by the quantum numbers $(\Delta;j_l,j_r;s; R)$ of its highest weight state under dilatations, Lorentz,  and $su(2)_R\times u(1)_R$. This state is created by a superconformal primary operator ${\cal O}$, defined by $[S_\alpha^a,\cO(0)]=[\overline S_{\dot\alpha}^{\,a},\cO(0)]=0$.

  \smallskip

An interesting class of superconformal primaries are the so-called chiral primary operators ${\cal O}_I$, annihilated by
\beq
[\overline Q_{\dot\alpha}^{\,a},\cO_I]=0\,,
\label{CPOc}
\eeq
together with the conjugate anti-chiral   primaries ${\overline \cO_{\bar I}}$
\beq
[Q_{\alpha}^a,\overline \cO_{\bar I}]=0\,.
\eeq
Unitarity of the SCFT and   the anticommutators of the $\cN=2$ superconformal algebra
 \begin{align}
 \{ Q_\alpha^a, S_\beta^b\}&= \epsilon_{\alpha\beta}\epsilon^{ab}\left(\Delta+{R\over 2}\right)+\epsilon^{ab} M_{\alpha\beta}+\epsilon_{\alpha\beta} J^{ab} \label{acone} \\[+2pt]
 \{ \overline Q_{\dot \alpha}^a, \overline S_{\dot \beta}^b\}&=\epsilon_{\dot\alpha\dot\beta}\epsilon^{ab}\left(\Delta-{R\over 2}\right)+\epsilon^{ab} M_{\dot\alpha\dot\beta}+\epsilon_{\dot\alpha\dot\beta} J^{ab}\label{actwo}\,
 \end{align}
 imply that\footnote{In our conventions $[R, Q_\alpha^a]=-Q_\alpha^a$\,.}
 \begin{align}
 {\cal O}_I:\  \Delta&={R\over 2}\,,\qquad\,\  \ \,  j_r=s=0~,\\[+4pt]
  \overline \cO_{\bar I}:\  \Delta&=-{R\over 2}\,,\qquad  j_l=s=0\,.
 \end{align}
Therefore, a  chiral  primary must transform as a scalar under $su(2)_R$ and its dimension   
 is completely determined  by its $u(1)_R$ charge $R$. A priori, a chiral primary can carry Lorentz spin $(j_l,0)$.  
 However, for SCFTs that admit a Lagrangian description somewhere in their conformal manifold, one can easily show that all chiral primaries must be Lorentz scalars, so that $j_r=j_l=0$. Furthermore, no example of a chiral primary with spin has been found to date  in  
  non-Lagrangian theories.\footnote{If spinning chiral primaries existed,  they  would be visible in the   superconformal index~\cite{Kinney:2005ej}. We are grateful to Leonardo Rastelli for a discussion.} See \cite{Buican:2014qla} for a further discussion about spinning chiral primaries. Henceforth, we only discuss  chiral primary operators that are Lorentz scalars. 
 These chiral primary operators parametrize the Coulomb branch of vacua of the SCFT, where $su(2)_R$ is preserved and $u(1)_R$ is spontaneously broken.
  
   \smallskip

  A chiral primary operator  of dimension $\Delta$ can be realized as the bottom component of an $\cN=2$ chiral  superfield $\mathcal{O}$ of Weyl weight $\Delta$ (we denote the superfield by the same symbol as the bottom component). This superfield  is annihilated by the four-dimensional $\cN=2$ right-handed superspace derivatives 
  \beq
  \overline D_{\dot \alpha}^{\,a} \mathcal{O} =0\,.
  \eeq
 The spacetime integral of the top component of a chiral   superfield   with $\Delta=2$, denoted by $C$,  defines an $\cN=2$ superconformal invariant, constructed by integrating the chiral superfield over the chiral half of the $\cN=2$ superspace\footnote{See Appendix \ref{ap:deform} for some details about the component structure of a chiral multiplet.}
 \beq
  \int d^4x\, d^4\theta\, \mathcal{O} =\int d^4x\, C \,.
 \eeq
   Therefore, chiral primary operators with $\Delta=2$, which we denote by $\cO_i$, give rise to exactly marginal operators, $C_i$.
   Geometrically, the $C_i$ can be viewed    as tangent vectors to the conformal manifold. 
      \smallskip

An important property of  chiral  primary operators in $\cN=2$ SCFTs is that they cannot disappear from the spectrum as we explore the conformal manifold. This is because the short representation  of the $\cN=2$ superconformal algebra built out of a 
chiral primary  highest weight cannot combine (at a generic point) with any other   multiplet of the $\cN=2$ superconformal algebra to become a long multiplet (see \cite{Dolan:2002zh} for the list of possible multiplet recombinations).
   \smallskip

   While chiral primary operators cannot disappear,   they  can mix when transported around the conformal manifold. Thus,  chiral primary operators can  be described as sections  of a holomorphic vector bundle over the conformal manifold~\cite{Papadodimas:2009eu}. The connection captures the operator mixing~\cite{Seiberg:1988pf,Kutasov:1988xb}.\footnote{Operator mixing is nontrivial when  the curvature of the connection is non-vanishing.  In $\cN=4$ super-Yang-Mills and for   the Higgs branch operators in $\cN=2$ SCFTs the situation is rather simple due to the fact that the corresponding curvatures vanish. This is however  not the case for chiral primaries (which we study in this paper) in    $\cN=2$ SCFTs.}

   \smallskip

  The operator product expansion (OPE) of chiral primary operators  is non-singular since  singular terms in the OPE would necessarily violate the unitarity bound $\Delta\geq R/2$. Therefore, 
   chiral primary operators furnish a ring, the chiral ring
  \beq
  \label{OPE}
  \cO_I(x) \cO_J(0)=\sum_K C_{IJ}^K \cO_K(0)+\ldots\,,
  \eeq
  where $\ldots$ denote $\overline Q$-exact terms.
  The multiplicative operation in this commutative ring is the CFT OPE. It is believed that for $\cN=2$ SCFTs the chiral ring is {\it freely generated}, that is, there exists a finite-dimensional basis of chiral operators such that any element of the chiral ring has a unique representation as a  polynomial in the basis elements. For Lagrangian $\CN=2$ theories, it is easy to show that indeed the chiral ring is freely generated.  The number of generators of the chiral ring is the dimension of the Coulomb branch of the SCFT.
     \smallskip

For a freely generated ring, we can always ``diagonalize" the product structure in the ring such that 
\begin{equation}
\cO_{I}(x)\cO_{J}(0)=\cO_{I}\cO_{J}(0)+...\label{normalizationconvention}\,,
\end{equation}
so that the matrix $\left(C_I\right)_J^K$ in (\ref{OPE}) has a single nonzero entry for each row.  While in this basis the ring structure constants are trivialized, the two-point functions of chiral primaries with anti-chiral primaries are nontrivial functions of the coupling constants
\begin{equation}\label{twopointchiral}
\vev{\cO_I(x)\overline{\cO}_{\bar J}(0)}_{\{\tau^i,\bar \tau^{\bar i} \}  }=\frac{G_{I\bar{J}}(\tau^i,\bar \tau^{\bar i})   }{|x|^{2\Delta_I}}\delta_{\Delta_I\Delta_{\bar J}}\,.
\end{equation} 
The metric $G_{I\bar{J}}$ defined by the two-point functions~\eqref{twopointchiral} is a Hermitian metric on the vector bundle. In this basis, the chiral ring data is captured by the Hermitian metric $G_{I\bar{J}}$.

 \smallskip

For completeness we would like to remind that  $\cN=2$ SCFTs contain another class of  half-supersymmetric  superconformal primary operators, ${\cal H}_I$. These are annihilated by supercharges of both chiralities\footnote{The two conditions are compatible since $\{Q_{\alpha}^{\,1},\overline Q_{\dot\alpha}^{\,1}  \}=0$.}
\beq
[Q_{\alpha}^{\,1},{\cal H}_I]=[\overline Q_{\dot\alpha}^{\,1},{\cal H}_I]=0\,.
\label{CPOh}
\eeq
Unitarity and the anticommutation relations \rf{acone}\rf{actwo} imply that ${\cal H}_I$ obey
\beq
\Delta= 2 s\,,\qquad\,\  \ \,  j_l=j_r=R=0\,. 
\eeq
Thus, these operators are Lorentz scalars, have vanishing $u(1)_R$ charge and    the conformal dimension is completely determined  in terms of the $su(2)_R$ isospin $s$. Furthermore,  they are highest weight of $su(2)_R$. The operators  ${\cal H}_I$ form a ring under the OPE, but unlike the chiral ring, this one  is not freely generated. The operators in this ring parametrize the Higgs branch of vacua of the SCFT, where $u(1)_R$ is unbroken and $su(2)_R$ is spontaneously broken. 

 \smallskip

The representations of the $\cN=2$ superconformal algebra with highest weight ${\cal H}_I$ with $s>3/2$ can recombine with other short multiplets  of the $\cN=2$ superconformal algebra to become a long representation.\footnote{We would like to thank Leonardo Rastelli for discussions about multiplet recombination.}
 The operators which do not recombine can  be described as sections  of  a vector bundle over the conformal manifold.  The curvature of this connection is vanishing. The  ring data associated to these operators is independent of   the exactly marginal couplings    \cite{Beem:2013sza, Baggio:2012rr}. This is unlike the chiral ring data which we study in this paper, where there is a nontrivial dependence on the exactly marginal couplings.

 \smallskip

In this paper we  relate the chiral ring data, $G _{I\bar{J}}$, of arbitrary $\cN=2$ SCFTs  admitting a Lagrangian description somewhere in the conformal manifold
to a certain partition function of the SCFT on $S^4$. This partition function, in turn,    can be computed exactly by supersymmetric localization.  More precisely, one can determine the $S^4$ partition function to all orders in perturbation theory and in some, but not all, cases also the instanton corrections. We will discuss this in detail in the main body of the paper.

\smallskip
From the chiral ring of $\cN=2$ SCFTs we can obtain  all of the  so-called extremal correlators
\begin{equation}\label{extremal}
\vev{\cO_{I_1}(x_1)\cO_{I_2}(x_2)...\cO_{I_n}(x_n){\overline \cO}_{\bar J}(y)} 
\end{equation}
everywhere on the conformal manifold, 
where by the $u(1)_R$ selection rule
\beq
\Delta_{I_1}+\Delta_{I_2}+...+\Delta_{I_n}=\Delta_{\bar J}\,.
\eeq
These correlators are, in general, non-holomorphic functions of $\tau^i,\bar \tau^{\bar i}$. Since there is only one anti-chiral operator in~\eqref{extremal}, these correlators are, in some sense, the simplest non-holomorphic local observables in the theory.
\smallskip

Let us now demonstrate that the extremal correlators~\eqref{extremal} can be obtained from the chiral ring data.   Without loss of generality we can put the operator ${\overline \cO}_{\bar J}$ at infinity by writing as   usual ${\overline \cO}_{\bar J}(\infty)\equiv \lim_{y\rightarrow\infty}y^{2\Delta_J} {\overline \cO}_{\bar J}(y)$. The next step is to observe that 
$\vev{\cO_{I_1}(x_1)\cO_{I_2}(x_2)...\cO_{I_n}(x_n){\overline \cO}_{\bar J}(\infty)}$ is independent of the coordinates $x_i$. 
One proves this by differentiating the   correlator with respect to the position of the $k$-th chiral primary and noting that
\beq
{\partial \over  \partial{x_k^{\alpha\dot\alpha}}} \cO_{I_k}(x_k)\propto \epsilon_{ab}\{\overline Q_{\dot \alpha}^a, [Q_{\alpha}^b,  \cO_{I_k}]\}\,.
 \eeq
By the supersymmetry Ward identity, we can   let  $\overline Q_{\dot \alpha}^a$ act on the rest of the operators. Using that $[\overline Q_{\dot \alpha}^a,\cO_{I}]=0$ and that $\overline Q_{\dot \alpha}^a$  acting on  ${\overline \cO}_{\bar J}(y)$ yields a correlator that decays as $y^{-2\Delta_J-1}$ completes the proof.  
 Therefore,  since $\vev{\cO_{I_1}(x_1)\cO_{I_2}(x_2)...\cO_{I_n}(x_n){\overline \cO}_{\bar J}(\infty)}$ is independent of the coordinates $x_i$ we can bring all the chiral primaries on top of each other and repeatedly use the OPE~\eqref{normalizationconvention} to reduce any extremal  correlation function to a two-point function in the chiral ring.  Then, if we know $G_{I\bar J}(\tau^i,\bar \tau^{\bar i})$ for all the $I,\bar J$, we are done.
 \smallskip

In the special case of maximally supersymmetric Yang-Mills theory ($\mathcal{N}=4$), extremal correlators have played an important role in the context of the AdS/CFT correspondence.
Indeed, it was   conjectured in~\cite{Lee:1998bxa,DHoker:1999ea,Intriligator:1998ig} that   extremal correlators can be computed exactly just from their tree-level diagrams, which allowed a comparison with supergravity. See \cite{Baggio:2012rr} for a  field theory proof of these nonrenormalization theorems in $\cN=4$ using Ward identities.  
 \smallskip

We will see that in general $\mathcal{N}=2$ theories there are both perturbative and non-perturbative corrections to extremal correlators.

\subsection{Subtle Aspects of Conformal Field Theories on $S^4$}

In this subsection our discussion pertains to general CFTs (i.e. not necessarily supersymmetric ones) in four dimensions. 
We can start from the CFT in flat space deformed by sources $\lambda^I(x)$ that couple to all the scalar primary operators $O_I(x)$
$$\int d^4 x \sum_I \lambda^I(x) O_I(x)~.$$
From the partition function $$Z[\bR^4](\lambda^I(x))$$ one can compute all the $n$-point functions of the scalar primary operators.
For example, it follows trivially that the one-point functions of all the operators other than the unit operator vanish. 

\smallskip

In order to define the theory on $S^4$, one needs to specify various additional contact terms. This is in spite of the fact that $S^4$ is conformally flat. The simplest example of the sort of subtleties that arise is the following: if there is an operator $O_0$ with $\Delta_0\in 2\mathbb{N}$, then we can add to the action the local counterterm\footnote{From now on $R$ will denote the Ricci scalar of the background metric and should not be confused with the $u(1)_R$ charge.}
\beq
\alpha \int d^4x\sqrt{g}\, \lambda_0\, R^{\Delta_0/2}\,\mathds{1}~,
\label{counter mix}
\eeq
with $R$ being the Ricci scalar (more generally, it could be a combination of Riemann tensors). Unlike separated-points correlation functions in flat space, this term depends on the scheme. As a result, the one-point function of $O_0$ on $S^4$ is scheme dependent  
$\langle O_0\rangle \sim \alpha r^{-\Delta_0}$, with $r$ being the radius of the sphere. $\alpha=0$ is obviously a preferred scheme, but it is not guaranteed that a given definition of the theory (say, by some RG flow) corresponds to this scheme.

\smallskip

Importantly for our analysis later, we can interpret $\alpha \int d^4x\sqrt{g}\, \lambda_0 R^{\Delta_0/2}\,\mathds{1}$ as a scheme-dependent operator mixing between $O_0$ and the unit operator $\mathds{1}$. This mixing can arise only in curved space, such as $S^4$. More generally, in curved space, the source for an operator $O_{\Delta_0}$  can have scheme-dependent non-minimal couplings to lower-dimensional operators due to nontrivial background fields, such as the curvature of space. This is only possible if the operators' dimensions differ by an even integer. 
These give rise to scheme-dependent operator mixing with all the operators of lower dimension in jumps by two units
\beq\label{mixing}O_{\Delta_0}\rightarrow O_{\Delta_0}+\alpha_1 R\, O_{\Delta_0-2}+\alpha_2 R^2\, O_{\Delta_0-4}+\cdots+\alpha_{\Delta_0/2}  R^{\Delta_0/2}\, \mathds{1} ~.
\eeq
If the CFT has exactly marginal couplings $\lambda^i$, then the coefficients $\alpha_k$ can depend on them. From the point of view of the CFT in $\bR^4$, the terms in~\eqref{mixing} induce contact terms between $O_{\Delta_0}$ and the energy-momentum tensor. These contact terms can be chosen at will according to the renormalization scheme. But once the theory is put  on $S^4$, these contact terms translate  to  operator mixing.

\smallskip

The conclusion from this discussion is that even for primary operators in a  CFT, the transition from $\bR^4$ to $S^4$ is nontrivial. One has to handle the possible operator mixing that is induced by various contact terms.

%%%%%%%%%%%%%%%%%%%%%%%%%%%%%%%%%%%%%%%%%%%%%%%%%%%%%
 
 \section{The Chiral Ring in 4d $\cN=2$ SCFTs and $S^4$}
 \label{sec:ring}
 
 In this section we explain how the chiral ring and   the extremal correlators \rf{extremalo} of an   $\cN=2$ SCFT  can be computed everywhere on the conformal manifold. Near a weakly coupled point on the conformal manifold, the answer can be in principle expanded into a perturbative series in the exactly marginal couplings $\tau^i,\bar \tau^{\bar i}$ dressed by an infinite sequence of instanton corrections. The key ingredient in obtaining the exact chiral ring data  is the relation we establish below with a   partition function on $S^4$. The $S^4$ partition function is of a suitable deformation of the    $\cN=2$ SCFT\@. For some theories,    the partition function  can      be explicitly evaluated by supersymmetric localization   using  formulae already available in the literature.

 \subsection{Placing the Deformed Theory on $S^4$}
 
 We are interested in studying the Lagrangian of an $\cN=2$  SCFT deformed by the top component of a  chiral multiplet  corresponding  to an arbitrary chiral primary operator $\cO$, which we denote by  $C$. This is done by adding to the Lagrangian in $\bR^4$ the following term\footnote{We change the normalization of the deformation by a factor of $1/32$ with respect to~\cite{Gerchkovitz:2014gta,Gomis:2014woa} in order to make formulae below simpler. In this normalization, the coefficient multiplying $K$ in equation  \rf{PartitionFun} should be $1/(2^{12}\times 3)$.} \begin{equation}\label{deform} -\frac{1}{32\pi^2}\tau_\cO\int d^4\theta \ \cO+\hbox{c.c.} =-\frac{1}{32\pi^2}\tau_\cO \, C   +\hbox{c.c.}\end{equation}
 If $\Delta(\cO)\neq 2$, this deformation  breaks the conformal symmetry as well as the $u(1)_R$ symmetry, while it preserves $su(2)_R$ and the $\cN=2$  super-Poincar\'e symmetry.
 If $\Delta(\cO) =2$ then the full $su(2,2|2)$ superconformal symmetry is preserved.
\smallskip
 
We will   show that the deformed SCFT  can be placed on $S^4$ while preserving $osp(2|4)$,  the supersymmetry algebra of the most general massive $\cN=2$ theory on $S^4$. The $so(2)_R\subset osp(2|4)$ is the Cartan generator of $su(2)_R$, and $sp(4)$ is the isometry of $S^4$.

 \smallskip

 We now explicitly construct the deformed SCFT on $S^4$. Placing the theory on $S^4$ requires deforming the flat space   expression~(\ref{deform}) by specific $1/r$ and $1/r^2$ terms, where $r$ is the radius of $S^4$, as in \cite{Festuccia:2011ws}.  The  deformed Lagrangian on $S^4$ can be derived by promoting  the coupling  $\tau_\cO$ in \rf{deform} to  a supersymmetric background chiral multiplet of Weyl weight $2- \Delta(\cO)$. The $osp(2|4)$ invariant   Lagrangian on $S^4$ is   constructed by deforming the SCFT with  the modified top component\footnote{Here $\tau_{1,2,3}^{ij}$ are the charge conjugated Pauli matrices defined as $\tau_p^{ij} \equiv \{i\sigma_3, -\mathds{1}_{2 \times 2}, -i\sigma_1\} =: \tau_{pij}^*$.} (see Appendix \ref{ap:deform})\vspace{+2pt}
  \beq
 \label{deformsphere}
\cC(x)\equiv  C(x)+2{(\Delta(\cO)-2)(\Delta(\cO)-3)\over r^2} \cO(x)- i {(\Delta(\cO)-2) \over r} \tau_1^{ij}B_{ij}(x)\,,
 \eeq
 where $B_{ij}$ is a middle component of the chiral multiplet $\cO$ (see Appendix \ref{ap:deform} for details of chiral multiplet components). Indeed, if we   add to the action of the SCFT on $S^4$ the deformation  $-\frac{\tau_{\cO}}{32\pi^2}\int d^4x \sqrt {g}\, \cC(x)+ \text{c.c.} $,   the   $osp(2|4)$ supersymmetry on $S^4$ is preserved.   In   superspace formalism, the sphere deformation \rf{deformsphere} is given by the following F-term
 \beq
-\frac{1}{32\pi^2} \int d^4x\int d^4\theta\,{\cal E}\, \tau_\cO\, \cO\,,
 \label{superspace}
 \eeq
 where ${\cal E}$ is the $\cN=2$ chiral density.
 
 \smallskip
 Note that for an exactly marginal deformation, which descends from    a chiral primary with $\Delta(\cO)=2$, there are no $1/r$ and  $1/r^2$ corrections in (\ref{deformsphere}). 
 
 \subsection{Chiral Primary Correlators from the Deformed Partition Function}

 We denote the partition function on $S^4$ of the deformed $\CN=2$ SCFT by
 \beq\label{deformedp}
Z[S^4](\tau^i,\bar\tau^{\bar i}\,; \tau^A,  \bar{\tau}^{\bar A})\,.
\eeq
 $\tau^A$ are the couplings associated to  chiral  ring generators  $\cO_A$ with $\Delta\neq 2$.
 We recall that $\tau^i$  are the couplings associated to the chiral primary operators   with $\Delta=2$, which are also chiral ring generators,    from which the exactly marginal operators are constructed. 
 
 \smallskip

 We can now study derivatives of the $S^4$ partition function with respect to the sources $\tau^I$ and $\bar{\tau}^{\bar J}$ where $\{\tau^I\}=\{\tau^i\}\cup\{\tau^A\}$. We consider first the 
  normalized second derivative  
\begin{equation}\label{second}{1\over Z[S^4](\tau^i,\bar\tau^{\bar i})} \partial_{\tau^I}\partial_{\bar{\tau}^{\bar I}} Z[S^4](\tau^i,\bar\tau^{\bar i}\,; \tau^A,  \bar{\tau}^{\bar A}) \biggr|_{\tau^A=\bar{\tau}^{\bar A}=0} =\left(\frac{1}{32\pi^2}\right)^2\int d^4x\sqrt{g(x)}\int d^4y\sqrt{g(y)} \,\langle \cC_I(x) \overline\cC_{\bar I}(y)\rangle_{S^4}\,.
 \end{equation}
This yields the integrated two-point function of the operator $\cC_I$  and $\overline \cC_{\bar I}$ in  \rf{deformsphere} on $S^4$. The integrated correlator  is  ultraviolet divergent, for example, due to the appearance of the unit operator in the OPE of $\cC_I$ and $\overline \cC_{\bar I}$, and must be regularized and renormalized. 

\smallskip

If we were to ignore supersymmetry for a moment, and if the sum of the dimensions of $\cC_I$ and $\overline \cC_{\bar I}$ were an even integer, the integrated correlation function \rf{second} would be ambiguous 
due to the local counterterm  
\beq\label{ct}
\int d^4x \sqrt{g}\, \tau^I\, \bar{\tau}^{\bar I}\, \cF(\tau^i,\bar\tau^{\bar i}) R^{\left(\Delta(\cO_I)+\Delta(\overline \cO_{\bar I})\right)/2}\,~,
\eeq
which shifts the result \rf{second} by an arbitrary function $\cF(\tau^i,\bar\tau^{\bar i})$.

\smallskip

Interestingly, in $\cN=2$ supersymmetric theories there is a unique way to regularize the divergences as $x\rightarrow y$ in~\eqref{second}. 
In other words, there is a unique way to regularize  the singularity $x\rightarrow y$ in a way consistent with  $\cN=2$ supersymmetry.  
There are two equivalent ways to understand this fact:

\begin{enumerate}
\item Using a   supersymmetry Ward identity on $S^4$ one can prove, extending the analysis in~\cite{Gomis:2014woa}, that (see Appendix~\ref{ap:ward} for the proof):
\begin{equation}\label{Ward}\int d^4x\sqrt{g(x)}\int d^4y\sqrt{g(y)} \,\langle \cC_I(x) \overline\cC_{\bar I}(y)\rangle_{S^4} = (32 \pi^2 r^2)^2\langle \cO_I(N)\overline \cO_{\bar I}(S)\rangle_{{S}^4}\,. \end{equation}
Therefore, a supersymmetric Ward identity shows that the supersymmetrically renormalized integrated correlation function of $\cC_I$ and $\overline\cC_{\bar I}$ in (\ref{deformsphere}) equals  the two-point function of the associated chiral primary $\cO_I$ at the North Pole of $S^4$ and of  the anti-chiral primary $\overline \cO_{\bar I}$ at the South Pole. 

\item  In a supersymmetric regularization, the counterterms~\eqref{ct} should  be  $\cN=2$ supergravity invariants. This restricts the allowed counterterms. Since $\tau^I$  and $\bar{\tau}^{\bar I}$ are embedded in a background  $\cN=2$ chiral and anti-chiral multiplet respectively, the counterterms that can lead to ambiguities in \rf{second} must be   D-term counterterms. Therefore, potential ambiguities can at best arise from superspace  integrals over all superspace $(\int d^4\theta\, d^4\bar\theta\,\cdot)$. But all the D-terms  vanish on supersymmetric backgrounds \cite{Butter:2014iwa}.\footnote{This result can be   derived 
 	by expressing the D-term invariants as   F-term invariants, constructed from a chiral integral  over half of the superspace $(\int d^4\theta\,\cdot)$  
 	using the chiral projector operator $\bar\Delta$  (see \cite{Kuzenko:2008ry} for details)
 	\beq
 	\int d^4x \int d^4\theta\, d^4\bar\theta\,E\, \cdot= \int d^4x \int d^4\theta\,{\cal E}\, \bar\Delta\,\cdot\,,
 	\eeq
 	where $E$ is the Berezinian and $\cE$ the chiral density of $\cN=2$ supergravity.
 	Since all terms in   $ \bar\Delta$ for $\cN=2$ supergravity are built out of the superspace derivatives $\overline D_{\dot \alpha}^{\,a}$ and $D_{\alpha}^{\,a}$
 	and supersymmetric configurations are annihilated by  $\overline D_{\dot \alpha}^{\,a}$ and $D_{\alpha}^{\,a}$, it follows that all D-terms vanish on supersymmetric backgrounds.  Since   $S^4$ is a supersymmetric background of a certain off-shell $\cN=2$ Poincar\'e supergravity theory~\cite{Gomis:2014woa} and the coupling constants, $(\tau^I, \bar{\tau}^{\bar I})$, are supersymmetric backgrounds of a chiral multiplet with the appropriate Weyl weight, all D-term counterterms automatically vanish.
 	This is to be contrasted with the chiral projector in e.g. 4d $\cN=1$ old minimal supergravity, where   $\bar\Delta=\bar D^2-8R$, and $R$ is a chiral superfield whose bottom component is the auxiliary field of old minimal supergravity. However, the situation in new minimal $\cN=1$ supergravity is rather similar to our present case~\cite{Assel:2014tba}.   
 	 We would like to thank Daniel Butter for helpful  discussions.}  Therefore, the singularity $x\rightarrow y$ in \rf{second} is regularized in a universal fashion.

\end{enumerate}
In summary, the two-point function of two arbitrary operators  in the chiral ring on $S^4$  can be obtained from the partition function of the deformed SCFT on $S^4$.  The relation between $S^4$ and $\bR^4$ correlation functions is not entirely straightforward, though. We will discuss this soon, after we review some properties of these four-sphere partition functions.

\subsection{The Deformed Partition Function on $S^4$}
\label{sec:defpart}
 
In the previous section we  showed   that an  $\cN=2$ SCFT on $S^4$ can be deformed with operators  that are descendants of operators in the chiral ring while preserving the $osp(2|4)$ symmetry of $S^4$. By adapting Pestun's  localization computation of the partition function of undeformed $\cN=2$ theories~\cite{Pestun:2007rz}, we can find the exact matrix integral representation for the partition function of the deformed SCFT on $S^4$.
\smallskip

We can localize the deformed partition function using the same supercharge ${\cal Q}$ in $osp(2|4)$ and $\cQ$-exact deformation term used in~\cite{Pestun:2007rz}. This supercharge obeys
\beq
\cQ^2=J^L_3+R\,,
\eeq
where $J^L_3$ is the Cartan generator of the $su(2)_L\subset sp(4)$ selfdual rotations on $S^4$ and $R$ is the Cartan generator of the $su(2)_R$ R-symmetry. This implies that the partition function localizes to the two fixed points of $J^L_3$ on $S^4$, that define the North and South Poles of $S^4$. Near the poles, the action of the deformed $\cN=2$ SCFT on $S^4$ approaches the action of the deformed $\cN=2$ SCFT in the $\Omega$-background~\cite{Nekrasov:2002qd}.
\smallskip

The deformed partition function on $S^4$ therefore localizes to the following matrix integral\footnote{$Z[S^4](\tau^i,\bar\tau^{\bar i},\tau^A,\bar \tau^{\bar A})$ should be thought of as a generating functional of correlators of chiral primary operators. We do not need to worry about its convergence properties at finite $\tau^{A}$.} 
\beq
Z[S^4](\tau^i,\bar\tau^{\bar i},\tau^{A},\bar \tau^{\bar A})=\int_{\mathfrak{t}} da\, \Delta(a)\, \left| Z_{\Omega}(a, \tau^i,\tau^{A})\right|^2\,.
\label{deformint}
\eeq 
As above, $\tau^{A}$ refers to the couplings associated  to the chiral ring generators with $\Delta \neq 2$.  In Lagrangian theories, the $\tau^{A}$ correspond to the higher Casimirs of the gauge group while $\tau^i$ to the quadratic Casimirs.  The matrix integral is over the Cartan subalgebra $\mathfrak{t}$ of the gauge group $G$ of the SCFT and $\Delta(a)$ is the associated Vandermonde determinant. $Z_{\Omega}(a, \tau^i,\tau^{A})$ is the partition function of the deformed SCFT in the $\Omega$-background  evaluated with equivariant rotation parameters $\varepsilon_1=\varepsilon_2=1/r$ and real equivariant parameters $a$ for the action of $G$.  From now on  we set $r=1$.  $Z_{\Omega}(a, \tau^i,\tau^{A})$ can, in turn,  be computed by supersymmetric localization, and takes the following form
\beq
Z_{\Omega}(a, \tau^i,\tau^{A})=Z_{\Omega,\text{cl}}(a, \tau^i,\tau^{A})\cdot Z_{\Omega,\text{loop}}(a)\cdot Z_{\Omega,\text{inst}}(a,\tau^i,\tau^{A})\,.
\label{productp}
\eeq 
The classical contribution for gauge group\footnote{It is trivial to extend this to any simple Lie group $G$. Then  $A$ takes values in the set of orders of the higher Casimirs of $G$. The formula   easily   extends when $G$ is product of simple gauge group factors, each giving rise to an exactly marginal deformation and a set of higher Casimir couplings.} $G=SU(N)$ is
 
\beq
Z_{\Omega,\text{cl}}(a, \tau,\tau^{A})=\exp{\left[i \pi   \tau\, \hbox{Tr} a^2+ i \sum_{A=3}^N \pi^{A/2}{\tau^A}\hbox{Tr} a^A\right]}\;.
\eeq
The one-loop determinant contribution is the same as in~\cite{Pestun:2007rz}, as it arises from the   ${\cal Q}$-exact deformation term
\beq
  |Z_{\Omega,\text{loop}}(a)|^2={\prod_{\alpha >0} H^2(i\alpha\cdot a)\over  \prod_{w\in {\bf r}} H(iw\cdot a)}\,,
\eeq
where $H(x)=G(1+x)G(1-x)$ and $G(x)$ is the Barnes double-gamma-function, which obeys $G(1+x)=\Gamma(x) G(x)$, with $\Gamma(z)$ being Euler's gamma-function. 
The numerator is the contribution of the vectormultiplet, governed by a product over the positive roots of the Lie algebra  of $G$.\footnote{The Vandermonde determinant in terms of the roots is $\Delta(a)=\prod_{\alpha>0} (\alpha\cdot a)^2$.} The denominator is the hypermultiplet contribution. The product is over  the weights of the representation ${\bf r}$ of $G\times G_F$, where $G_F$ is the flavor symmetry acting on the hypermultiplet.\footnote{We set the equivariant parameters for $G_F$, i.e. the mass parameters, to zero.}   

\smallskip

$Z_{\Omega,\text{inst}}(a,\tau^i,\tau^A)$ captures the  contribution of  point-like instantons to the path integral~\cite{Nekrasov:2002qd}. The fact that it depends on $\tau^A$  means that one cannot just evaluate the operator insertions on the saddle points of the undeformed SCFT. This is because the operators are inserted precisely where point-like instantons and anti-instantons are localized, thus changing the saddle points themselves.
The instanton partition function is generally  given by a series expansion over the instanton charge. Roughly speaking, the contribution at a given instanton charge is obtained by integrating a certain equivariant characteristic class of a vector bundle over the corresponding moduli space of instantons. Important subtleties arise because the moduli space of instantons has singularities, and the integrals must be properly defined. There is a canonical way of  defining the integrals over instanton moduli space  when the gauge group is $U(N)$. In this case, singularities in the moduli space are resolved by turning on noncommutativity (see e.g. \cite{Nekrasov:2002qd}). In general, it is an open problem   to compute $Z_{\Omega,\text{inst}}(a,\tau^i,\tau^A)$ for $SU(N)$ with $N>2$.   Solving this problem will have some applications for our study of extremal correlators, but one can make some significant mileage even before this problem is solved. 
In section 3 we study examples in which $Z_{\Omega,\text{inst}}(a,\tau^i,\tau^A)$ is known as well as some examples where it is not known, but one can still study the perturbative series.

\subsection{The Relation Between Correlators in $\bR^4$ and $S^4$}\label{relation}

As we have explained above, using the deformed  partition function on $S^4$    \rf{deformint} and  the Ward identity~\eqref{Ward}, we can calculate, in particular,  the two-point functions of arbitrary chiral primary operators on $S^4$
\beq\label{alltwo}  \langle \cO_I(N)\overline \cO_{\bar J}(S)\rangle_{{S}^4}\,. \eeq
In this section we explain how to obtain 
the  two-point functions of chiral primary operators in flat space~\eqref{twopointchiral} from  the explicit results of the correlation functions on $S^4$ .

\smallskip
As explained in subsection 1.3, in the dictionary between CFT sphere correlation functions and flat space correlation functions   one expects operator mixing~\eqref{mixing},  induced by the background fields. In fact, in $\cN=2$ SCFTs we already know that such mixing must take place from the formula~\eqref{PartitionFun}.
This formula shows that the one-point function 
$
\langle \cO_i(N)\rangle_{{S}^4}=\frac{1}{Z[S^4]}{\partial\over \partial \tau^i} Z[S^4] 
$,
 is non-vanishing. This is a special case of~\eqref{mixing} since it can be interpreted as mixing of $\cO_i$ with the identity operator $\mathds{1}$. This mixing with the identity operator can be interpreted, in turn,  as a conformal anomaly according to~\cite{Gomis:2015yaa}. 

\smallskip
In complete generality, we should allow a chiral primary operator $\cO_\Delta$ of dimension $\Delta$ to mix with lower dimensional chiral operators 
\beq\label{mixingSUSY}
\cO_{\Delta}  \longrightarrow \cO_{\Delta}  +\alpha_1(\tau^i,\bar \tau^{\bar i})R\,  \cO_{\Delta-2}+\alpha_2(\tau^i,\bar \tau^{\bar i})R^2\,  \cO_{\Delta-4}+\cdots~,
\eeq
and similarly for the anti-chiral operators. In~\eqref{mixingSUSY} $R^k$ stands schematically for some contraction of $k$ Riemann tensors evaluated on the sphere. Note that the   chiral operator $\cO_\Delta$ can only mix with other chiral operators, and not anti-chiral or the Higgs branch operators ${\cal H}_I$ discussed in section \ref{sec:cpodet}. 
Indeed, while chiral operators are supersymmetric at the North pole of $S^4$, neither anti-chiral operators nor ${\cal H}_I$ are supersymmetric there. Anti-chiral operators are supersymmetric, instead, at the South pole of $S^4$, while the Higgs branch operators cannot be inserted anywhere on $S^4$ while preserving supersymmetry (just as operators in a long representation of the superconformal algebra). Since operator mixing   is compatible with supersymmetry on $S^4$, chiral primary operators can only   mix among themselves, and analogously for anti-chiral operators. 
\smallskip
 
It is natural to conjecture that the mixing coefficient functions $\alpha_k(\tau^i,\bar \tau^{\bar i})$ are captured by some anomalies, in parallel with the origin of the mixing of $\cO_i$ with the identity operator. Operator mixing of the type in~\rf{mixingSUSY} can only occur when the theory has operators with integer-spaced dimensions. We can then expect that the there would be various type-$B$ ``resonance'' anomalies. See for example \cite{Deser:1976yx,Banados:2006de,Bzowski:2015pba}. These anomalies generalize the Zamolodchikov anomaly studied in~\cite{Gomis:2015yaa}, which is  responsible for the mixing of $\cO_i$ with the identity operator. It would be very nice to understand this structure better. 

\smallskip

Since the mixing functions $\alpha_k(\tau^i,\bar \tau^{\bar i})$ are expected to arise due to anomalies, they are expected to be universal. There is, however, a holomorphic ambiguity, which acts by
\beq
\alpha_k(\tau^i,\bar \tau^{\bar i})\rightarrow \alpha_k(\tau^i,\bar \tau^{\bar i})+ \cF_k(\tau^i)+\overline \cF_k({\bar\tau}^{\bar i})\,.\label{16}
\eeq
Of course, the holomorphic ambiguity is fixed when the renormalization scheme is fixed. These holomorphic ambiguities in operator mixing 
are due to $\cN=2$ supersymmetric counterterms.  A special case of this holomorphic counterterm is responsible for the ambiguous mixing of $\cO_i$ with the unit operator, which  was already constructed in~\cite{Gomis:2014woa,Gomis:2015yaa}. This counterterm is responsible for the K\"ahler ambiguity of the partition function of the SCFT on $S^4$~\eqref{PartitionFun}.
\smallskip

When mapping  the $S^4$ correlation functions to the correlation functions on $\bR^4$ we must deal with the operator mixing in~\eqref{mixingSUSY}. Let us first review how this is accomplished  for the special case  of chiral primaries of dimension 2, $\cO_i$. We recall that their descendants are the exactly marginal deformations that generate the conformal manifold of the SCFT. On $S^4$,  there is mixing of $\cO_i$ with the unit operator, as follows from~\eqref{PartitionFun}.  In this special case,  it is easy  to disentangle  the operator mixing:  we simply subtract disconnected pieces in $\langle \cO_i(N)  {\overline \cO}_{\bar j}(S)\rangle_{S^4}$ from the right hand side of~\eqref{Ward}. It is well known that this can be achieved by taking the logarithm of the sphere partition function (which indeed removes all the disconnected diagrams). After we have removed this mixing, we can straightforwardly relate the $\langle \cO_i(N)  {\overline \cO}_{\bar j}(S)\rangle_{S^4}$  two-point functions with their flat space counterparts $\langle \cO_i(0)  {\overline \cO}_{\bar j}(\infty)\rangle_{\bR^4}$, from which the metric is extracted. Therefore, the mixed second derivatives of the $\ln Z[S^4]$  with respect to the moduli $\tau^i, \bar \tau^{\bar i}$ compute the Zamolodchikov metric on the conformal manifold. This is precisely the statement captured by~\eqref{PartitionFun}. 
\smallskip

In more generality, for higher-dimensional chiral primaries, there can be nontrivial mixing with  all  the chiral primary operators of lower dimension, and taking the logarithm of the sphere partition would not suffice to remove operator mixing. 
In this case,  diagonalization of $\langle \cO_I(N)\overline \cO_{\bar J}(S)\rangle_{{S}^4}$ must be carried out, which can be implemented by a {\it Gram-Schmidt} procedure. This prescription  is  the appropriate generalization of the ideas leading to~\eqref{PartitionFun}. As we will see, this approach to computing flat space correlation function successfully reproduces many perturbative results while providing many new results, and it satisfies nontrivial all-orders consistency checks. We now summarize the explicit   algorithm   to determine the chiral ring data of an $\cN=2$ SCFT.

\subsection{Summary of the Algorithm}\label{summary}

We consider an $\cN=2$ SCFT with exactly marginal couplings $\tau^i,\bar \tau^{\bar i}$. 
The chiral ring is finitely generated and we take the generators to be $\phi_{\alpha}$, $\alpha=1,...,\frak{N}$, with $\frak{N}$ the number of generators. $\frak{N}$ is also the dimension of the Coulomb branch of the SCFT. We denote their dimensions by $\Delta(\phi_{\alpha})=\Delta_\alpha$. Every element in the chiral ring can be uniquely represented as a linear combination of 
\beq\label{representation}\cO_{n_1,...,n_{\frak{N}}} = \phi_1^{n_1}\phi_2^{n_2}...\phi_{\frak{N}}^{n_\frak{N}}~. \eeq
 
\smallskip
The  Lagrangian of the  SCFT is constructed from the ring generators with $\Delta=2$. 
We now deform the SCFT   using the chiral ring generators of $\Delta>2$, which we denote by  $\phi_A$ 
\beq \label{defo} S_{\text{SCFT}}\rightarrow S_{\text{SCFT}}- \frac{1}{32\pi^2} \int d^4x\, d^4\theta \, {\cal E} \, \sum_{A} \tau^{A} \phi_{A}+\hbox{c.c.}\eeq
This is  appropriately supersymmetrized on $S^4$, as explained in subsection 2.1.
The associated partition function \rf{deformint} is denoted by 
\beq\label{deformedPa}Z[S^4](\tau^i,\bar \tau^{\bar i} ; \tau^{A},\bar \tau^{\bar A} )~.\eeq

\smallskip

Our goal is to compute  the two-point functions in flat space 
$$\langle \cO_{n_1,...,n_{\frak{N}}} (0) \overline \cO_{n'_1,...,n'_{\frak{N}}} (\infty) \rangle_{\bR^4}\,.$$
These are possibly nonzero only if $\Delta\equiv\sum_{\alpha=1}^{\frak{N}} n_\alpha\Delta_\alpha= \sum_{\alpha=1}^{\frak{N}} n'_\alpha\Delta_\alpha $.  Given $Z[S^4](\tau^i,\bar \tau^{\bar i} ; \tau^A,\bar \tau^{\bar A} )$, we must first disentangle the operator mixing of 
$\cO_{n_1,...,n_{\frak{N}}}$ and   $\cO_{n'_1,...,n'_{\frak{N}}}$ on $S^4$ with the lower-dimensional chiral operators, as described in \eqref{mixingSUSY}. In order to do this, we implement the following procedure:

\begin{enumerate}
\item List  all chiral  operators $\cO_{m_1,...,m_{\frak{N}}}$ of dimension  
$\sum_{\alpha=1}^{\frak{N}} n_\alpha\Delta_\alpha-2$, $\sum_{\alpha=1}^{\frak{N}} n_\alpha\Delta_\alpha-4$ etc. We denote the number of    operators up to dimension $\Delta-2$ by $N_{\Delta-2}$. 

\item Compute the $N_{\Delta-2}+1$ dimensional matrix of two-point functions on the sphere $$\langle \cO_{m_1,...,m_{\frak{N}}} (N) \overline \cO_{m'_1,...,m'_{\frak{N}}} (S) \rangle_{S^4}\equiv M_{ m_1,...,m_{\frak{N}}   | m'_1,...,m'_{\frak{N}}  }$$ for all the  operators listed in the previous step and for the operator $\cO_{n_1,...,n_{\frak{N}}}$ in question. This Hermitian matrix is generally nonzero in all its entries. Do the same for the operator $\overline \cO_{n'_1,...,n'_{\frak{N}}}$.

\item From~\eqref{deformedPa} we  can extract  the matrix $M_{ m_1,...,m_{\frak{N}}   | m'_1,...,m'_{\frak{N}}  }$ by 
\beq\label{sphereMa}   M_{ m_1,...,m_{\frak{N}}   | m'_1,...,m'_{\frak{N}}  }= {1\over Z[S^4](\tau^i,\bar \tau^{\bar i}) }{\partial^{m_1}\over (\partial \tau^1)^{m_1}   }\cdots{   \partial^{m_{\frak{N}}}  \over (\partial \tau^{\frak{N}})^{m_{\frak{N}}}   }{\partial^{m'_1}\over (\partial\bar \tau^1)^{m'_1}   }\cdots{   \partial^{m'_{\frak{N}}}  \over (\partial \bar \tau^{\frak{N}})^{m'_{\frak{N}}}   }Z[S^4] \biggr|_{\tau^{A}=\bar\tau^{\bar A} =0   }  \eeq

\item The mixing of the operator   $\cO_{n_1,...,n_{\frak{N}}}$  on $S^4$ with lower-dimensional operators~\eqref{mixingSUSY} is encoded in $N_{\Delta-2}$ coefficients $\alpha_k(\tau^i,\bar\tau^{\bar i})$. These can be determined uniquely by demanding that the two-point function of $\cO_{n_1,...,n_{\frak{N}}}$ with each one of the $N_{\Delta-2}$ lower dimension operators vanishes. Do likewise for the operator $\overline \cO_{n'_1,...,n'_{\frak{N}}}$.

\item This  algorithm is equivalent to performing a Gram-Schmidt diagonalization procedure of the matrix $M_{ m_1,...,m_{\frak{N}}   | m'_1,...,m'_{\frak{N}}  }$. After completing this procedure for $\cO_{n_1,...,n_{\frak{N}}}$ and $ \cO_{n'_1,...,n'_{\frak{N}}}$, the two-point function  of   orthogonalized operators on $S^4$ are directly related to  $\langle \cO_{n_1,...,n_{\frak{N}}} (0) \overline \cO_{n'_1,...,n'_{\frak{N}}} (\infty) \rangle_{\bR^4}$.

\end{enumerate}
\medskip

Let us show that the formula~\eqref{flatcorrelaxxx} is a special case of the procedure outlined above. We are interested in the two-point functions of $\Delta=2$ chiral operators in $\bR^4$. Let us assume for notational simplicity that there is only one such $\Delta=2$ operator. The matrix of two-point functions on the sphere is therefore a $2\times 2$ matrix: 
\beq\label{22example} {1\over Z[S^4]}\left( \begin{matrix} Z[S^4] & \partial_\tau Z[S^4] \cr \partial_{\bar \tau} Z[S^4]& \partial_\tau\partial_{\bar \tau} Z[S^4] \end{matrix} \right)\,.\eeq
We perform the Gram-Schmidt procedure and find the norm of the corresponding non-trivial orthogonal vector. This is given by the determinant of~\eqref{22example}, namely, \beq\label{22examplei}\langle \cO(0) \overline \cO (\infty) \rangle_{\bR^4}\sim {1\over (Z[S^4])^2}\left(Z[S^4]\partial_\tau\partial_{\bar \tau} Z[S^4]-\partial_\tau Z[S^4]\partial_{\bar \tau} Z[S^4]\right)~.\eeq This combination coincides with $\partial_\tau\partial_{\bar \tau} \ln  Z[S^4]$.

\smallskip

 We now discuss several examples to further demonstrate the procedure and its various applications and consequences.

\section{Examples}
\label{sec:Toda}

\subsection{$SU(2)$ Gauge Group}

The first example we   consider is $\cN=2$ SCFTs with gauge group $SU(2)$.
The discussion in this subsection applies both to superconformal $SU(2)$ SQCD with 
  four fundamental hypermultiplets and   to $\cN=4$ $SU(2)$ super-Yang-Mills. 
\smallskip 

  The chiral ring in this case has one generator, $\phi_2=-4\pi i\Tr\varphi^2$, where $\varphi$ is the complex scalar in the vectormultiplet. Thus, the chiral ring operators  are given by 
\beq
\cO_n=(\phi_2)^n\;,\;\;\;n\in\mathbb{N}\;,\label{CPOsu2}
\eeq
with $\cO_0\equiv \mathds{1}$.  The chiral ring  OPE is  
\begin{equation}
\cO_{n}(x)\cO_{m}(0)=\cO_{n+m}(0)+...\;.
\end{equation}
Since in this case there is a single chiral primary with $\Delta=2$, the conformal manifold is one-complex-dimensional. In gauge theory terms, the   complex coordinate in the conformal manifold is  given  by the complexified gauge coupling $\tau = \frac{\theta}{2\pi} + i\frac{4\pi}{g^2}$, where $g$ is the Yang-Mills coupling and $\theta$ is the  theta angle.
\smallskip

We now study  the problem of computing all the flat space two-point functions~\rf{twopointchiral}
\beq
G_{2n}(\tau,\bar \tau)=\vev{\cO_n(0)\overline \cO_n(\infty)}_{\bR^4}~.\label{g2ndef}
\eeq
This determines the chiral ring data and the extremal correlators \rf{extremalo}. 
Obviously $G_0=1$, and it follows from~\eqref{flatcorrelaxxx} 
that 
\beq G_2=16\,\partial_\tau \partial_{\bar\tau} \ln Z[S^4]\;.\label{BoundaryCondition}
\eeq 
 Alternatively, this formula can be derived from our Gram-Schmidt procedure as in~\eqref{22examplei}. 
\smallskip

We now follow the algorithm described in the previous section, and   begin by studying 
the two-point functions of the operators (\ref{CPOsu2}) on $S^4$, $\vev{\cO_n(N)\overline \cO_m(S)}_{S^4}$. These two-point functions define an inner product on the chiral ring. As in~\eqref{sphereMa}, we express these two-point functions as derivatives of the sphere partition function\footnote{For  $SU(2)$ SQCD with 4 hypermultiplets in the fundamental representation one finds (see subsection \ref{sec:defpart})
\begin{equation}   \label{N2Z}
Z[S^4](\tau,\bar{\tau})=\int_{-\infty}^{\infty}da\, e^{-4\pi\, {\text{Im}}\tau \,a^2}(2a)^2\frac{H(2ia)H(-2ia)}{[H(ia)H(-ia)]^4} |Z_{\Omega,\text{inst}} (ia,\tau)|^2\;,
\end{equation}
 and $Z_{\Omega,\text{inst}}$ is Nekrasov's instanton partition function on the $\Omega$-background~\cite{Nekrasov:2002qd}.
By expanding the integrand in powers of $g^2$ we can compute $Z[S^4]$ to any order in perturbation theory, and we can also include instantons. In $SU(2)$ gauge theory with an adjoint hypermultiplet, i.e. $\mathcal{N}=4$ $SU(2)$ super-Yang-Mills, the result is much simpler 
\begin{equation} \label{N4} 
Z[S^4](\tau,\bar{\tau})=\int_{-\infty}^{\infty}da\, e^{-4\pi\, {\text{Im}}\tau \,a^2}(2a)^2~.  
\end{equation}
} 
\beq
\langle \cO_n(N) \overline \cO_m(S) \rangle_{S^4} = \frac{1}{Z[S^4]} \partial_\tau^n \partial_{\bar\tau}^m Z[S^4]\;. \label{cor1}
\eeq
The basis of operators $\{\cO_n\}$ is not orthogonal with respect to this inner product. We diagonalize the mixing by  carrying  out   the Gram-Schmidt construction in order to  find a basis $\{\cO_n\} \to \{\cO'_n\}$,
such that the new operators, given by  
\beq 
\cO'_n = \cO_n - \sum_{m=0}^{n-1} \frac{\langle \cO_n(N) \overline\cO'_m(S)\rangle}{\langle \cO'_m(N) \overline\cO'_m(S)\rangle} \cO'_m\;  \label{GS}
\eeq
are mutually orthogonal. 
The two-point functions on $S^4$ in this new basis can now be identified with the two-point functions in flat space $\vev{\cO_n(0)\overline \cO_n(\infty)}_{\bR^4}$. Therefore, 
\beq
\langle \cO'_n (N)\overline\cO'_m (S) \rangle_{S^4} \equiv \frac{1}{16^{n}} G_{2n}(\tau,\bar \tau)\delta_{nm}\;.\label{gdef}
\eeq

\smallskip

The Gram-Schmidt diagonalization procedure~\eqref{GS} 
  is   recursive, and    can be solved explicitly for arbitrary $n$. By virtue of~\eqref{cor1}, the  orthogonal vectors  can be expressed in terms of derivatives of $Z[S^4]$. Therefore, we can   express the chiral ring data $G_{2n}(\tau,\bar \tau)$ in terms of various derivatives of the $S^4$ partition function $Z[S^4]$. This  suggests, in turn, that   the various metrics $G_{2n}(\tau,\bar \tau)$ can be related by differential equations. We will now prove this. 

\smallskip
For the purpose of exhibiting the system of differential equations acting on the chiral ring data  it is useful to organize the two-point functions on $S^4$ in~\eqref{cor1} in an infinite dimensional matrix 
\begin{equation}
M_{m,n} = \langle \cO_m(N) \overline \cO_n(S) \rangle_{S^4}\;, \qquad m,n = 0, 1, \cdots. \label{aux1}
\end{equation} 
Let us denote by $M_{(n)}$ the upper-left $(n+1)\times(n+1)$ submatrix of $M$, and 
\begin{equation}
D_n \equiv\; \det M_{(n)}\;.
\end{equation}  
This submatrix captures the mixing of the operator $\cO_n$ with all operators of smaller dimension, i.e. $\Delta<2n$. Because the matrices that appear in the Gram-Schmidt procedure are triangular (operators can only mix with lower-dimensional operators), one can obtain   $G_{2n}(\tau,\bar\tau)$ in~\eqref{gdef} as a ratio of determinants
\beq
G_{2n}(\tau,\bar\tau)= 16^{n}\frac{D_n}{D_{n-1}}\;. \label{gD}
\eeq
In addition, we can prove that the determinant $D_n$ satisfies the differential equation\footnote{To prove this, we first write the derivative of $\ln D_n$ in terms of derivatives of $M_{(n)}$ as follows:
\beq
\partial_\tau \partial_{\bar\tau} \ln D_n = \Tr\left( M_{(n)}^{-1} \partial_\tau \partial_{\bar\tau} M_{(n)} - M_{(n)}^{-1} \partial_\tau M_{(n)} M_{(n)}^{-1} \partial_{\bar\tau} M_{(n)} \right). \label{dDM}
\eeq
Using (\ref{aux1}) and (\ref{cor1}), the derivatives of the components of $M$ can be written as:
\begin{subequations}
	\begin{align}
	\partial_\tau M_{i,j} =&\; M_{i+1,j} - M_{1,0}M_{i,j}, \\
	\partial_{\bar\tau} M_{i,j} =&\; M_{i,j+1} - M_{0,1} M_{i,j}, \\
	\partial_\tau \partial_{\bar\tau} M_{i,j} =&\; M_{i+1,j+1} - M_{1,0} M_{i,j+1} - M_{0,1} M_{i+1,j} + (2M_{1,0}M_{0,1} - M_{1,1}) M_{i,j}.
	\end{align} \label{dM}
\end{subequations}
Using these relations and noting that $D_1 = M_{1,1} - M_{1,0}M_{0,1}$, we arrive at
\beq
\partial_\tau \partial_{\bar\tau} \ln D_n = \left(M_{(n)}\right)_{n,n}^{-1}\left( M_{n+1,n+1} - \sum_{i,j=0}^n M_{n+1,i} \left(M_{(n)}\right)_{i,j}^{-1} M_{j,n+1} \right) - (n+1) D_1. \label{dDM2}
\eeq
Using Schur's complement lemma
\beq
M_{n+1,n+1} - \sum_{i,j=0}^n M_{n+1,i} \left(M_{(n)}\right)_{i,j}^{-1} M_{j,n+1} = \frac{D_{n+1}}{D_n}, \quad \mathrm{and} \quad \left(M_{(n)}\right)_{n,n}^{-1} = \frac{D_{n-1}}{D_n}\;,
\eeq
we obtain~\eqref{DED}.} 
\beq
\partial_\tau \partial_{\bar\tau} \ln D_n = \frac{D_{n+1} D_{n-1}}{D_n^2} - (n+1) D_1\,. \label{DED}
\eeq
\smallskip

Combining~\eqref{gD} and \eqref{DED} we find an equation directly for  the two-point functions $G_{2n}(\tau,\bar\tau)$
\beq
16\,\partial_\tau \partial_{\bar\tau} \ln G_{2n} = \frac{G_{2n+2}}{G_{2n}} - \frac{G_{2n}}{G_{2n-2}} - G_2\;,\;\;\;\;\;n=1,2...\;. \label{rec}
\eeq
Recall that $\{G_{2n}(\tau,\bar\tau)\}$ obey the following boundary conditions: $G_0=1$ and
 $G_2=16\,\partial_\tau \partial_{\bar\tau} \ln Z[S^4]$. By defining $G_{2n} \equiv 16^n\,e^{q_n - \ln Z[S^4]}$, the differential equation (\ref{rec}) can be cast into the form of the   semi-infinite Toda chain equation
\begin{equation}\label{toda}
\begin{aligned}
&\partial_\tau \partial_{\bar\tau} q_n = e^{q_{n+1}-q_n} - e^{q_n-q_{n-1}}\;,\qquad n = 1,2,\cdots\\
&\partial_\tau \partial_{\bar\tau}q_0=e^{q_1-q_0}\,.
\end{aligned}
\end{equation} 
Therefore, the chiral ring data is governed by a system of coupled oscillators with a prescribed dependence on $\tau,\bar\tau$ for the leftmost oscillator, that is $q_0=\ln Z[S^4]$. In this particle picture, 
we can think of ${\rm Im}\, \tau$ as physical time. Since ${\rm Re}\, \tau$ is compact, we can Fourier decompose in it and imagine that the lattice has two spatial dimensions.

\smallskip
We see that the Toda chain~\eqref{toda} arises essentially from the Gram-Schmidt procedure on $S^4$, with the ratio of some determinants~\eqref{aux1}--\eqref{gD} playing a central role.  This is in fact reminiscent of the way solutions to the semi-infinite Toda system are actually constructed in the integrability literature \cite{hirota1988solutios}. 
\smallskip

In~\cite{Baggio:2014sna}, the $tt^*$ equations of four-dimensional $\cN=2$ SCFTs in the holomorphic gauge were exploited to arrive at the same equations~\eqref{rec} (the $tt^*$ equations   do not provide the boundary condition~\eqref{BoundaryCondition}). This agreement with the $tt^*$ equations is therefore a nontrivial consistency check of our procedure. 
\smallskip

In Appendix \ref{ap:int} we show that the $tt^*$ equations of an arbitrary four-dimensional $\cN=2$ SCFT are integrable. They can be written as the flatness condition of a one-parameter family of connections  like the $tt^*$ equations of a    two-dimensional $(2,2)$ QFTs \cite{Cecotti:1991me}. The $tt^*$ equations  are governed by a Hitchin integrable system.
\smallskip

\subsubsection{$SU(2)$ with an Adjoint Hypermultiplet}

It is important to note that due to the simple form of the $S^4$ partition function given in~\eqref{N4}  for 
$\cN=2$ $SU(2)$ with an adjoint hypermultiplet, that is ${\cal N} = 4$ $SU(2)$ super-Yang-Mills, the partition function evaluates to 
 \begin{equation}
  Z_{S^4}[\tau, \bar \tau] = \frac{1}{4 \pi (\text{Im} \tau)^{3/2}} \,.
 \end{equation}
All the $G_{2n}(\tau,\bar\tau)$ coincide with their tree-level expressions
 \begin{equation}
  G_{2n}(\tau, \bar \tau) = G_{2n}^\text{tree} (\tau, \bar \tau) = \frac{(2n+1)!}{(\text{Im} \tau)^{2n}} = (2n+1)! \left( \frac{g^2}{4 \pi} \right)^{2n}  \,.
 \end{equation}
One can easily verify that indeed  these expressions obey the Toda equations \eqref{rec}.

\subsubsection{$SU(2)$ SQCD with Four Fundamental Hypermultiplets}

In the case of $SU(2)$ SQCD with four fundamental hypermultiplets, the $S^4$ partition function given in \eqref{N2Z} has quite a non-trivial dependence on $\text{Im} \tau = 4 \pi / g^2$, and the $G_{2n}(\tau, \bar \tau)$ receive both perturbative and non-perturbative corrections.  To reproduce this expansion, one can start with \eqref{N2Z} and expand  the instanton partition function   
 \begin{equation}
  Z_{\Omega, \text{inst}} (i a, \tau) = 1 + \frac 12 e^{2 \pi i \tau} (a^2 -3) + \cdots \,,
 \end{equation}
where the first term corresponds to the zero-instanton sector, the second term to the $1$-instanton sector, etc., as well as expand the functions $H$ in \rf{N2Z} at small $a$.  Order by order in these expansions, the integrals in $a$ are elementary.  The first few terms are
 \begin{equation}
  \begin{aligned}
   Z_{S^4}[\tau, \bar \tau] &= \frac{1}{4 \pi (\text{Im} \tau)^{3/2}}  \left[1 -  
   \frac{45 \zeta(3)}{16 \pi^2 (\text{Im} \tau)^2} + \frac{525 \zeta(5)}{64 \pi^3 (\text{Im} \tau)^3} + \cdots \right] \\[+2ex]
   &+ \frac{ e^{2 \pi i \tau}  + e^{-2 \pi i \bar \tau}}{8 \pi (\text{Im} \tau)^{3/2}}
    \left[ -3 + \frac{3}{8 \pi \text{Im} \tau} + \frac{135 \zeta(3)}{16 \pi^2 (\text{Im} \tau)^2 } + \cdots \right] + \cdots \,,
  \end{aligned}
 \end{equation}
where the first line contains the perturbative contributions and the second line contains the non-perturbative ones starting with the $1$-instanton result.  As we have explained,  this expression can be used to compute all the $G_{2n}$ in $SU(2)$ SQCD\@.

\begin{figure}[t]
\begin{center}
\includegraphics[width = 0.5\textwidth]{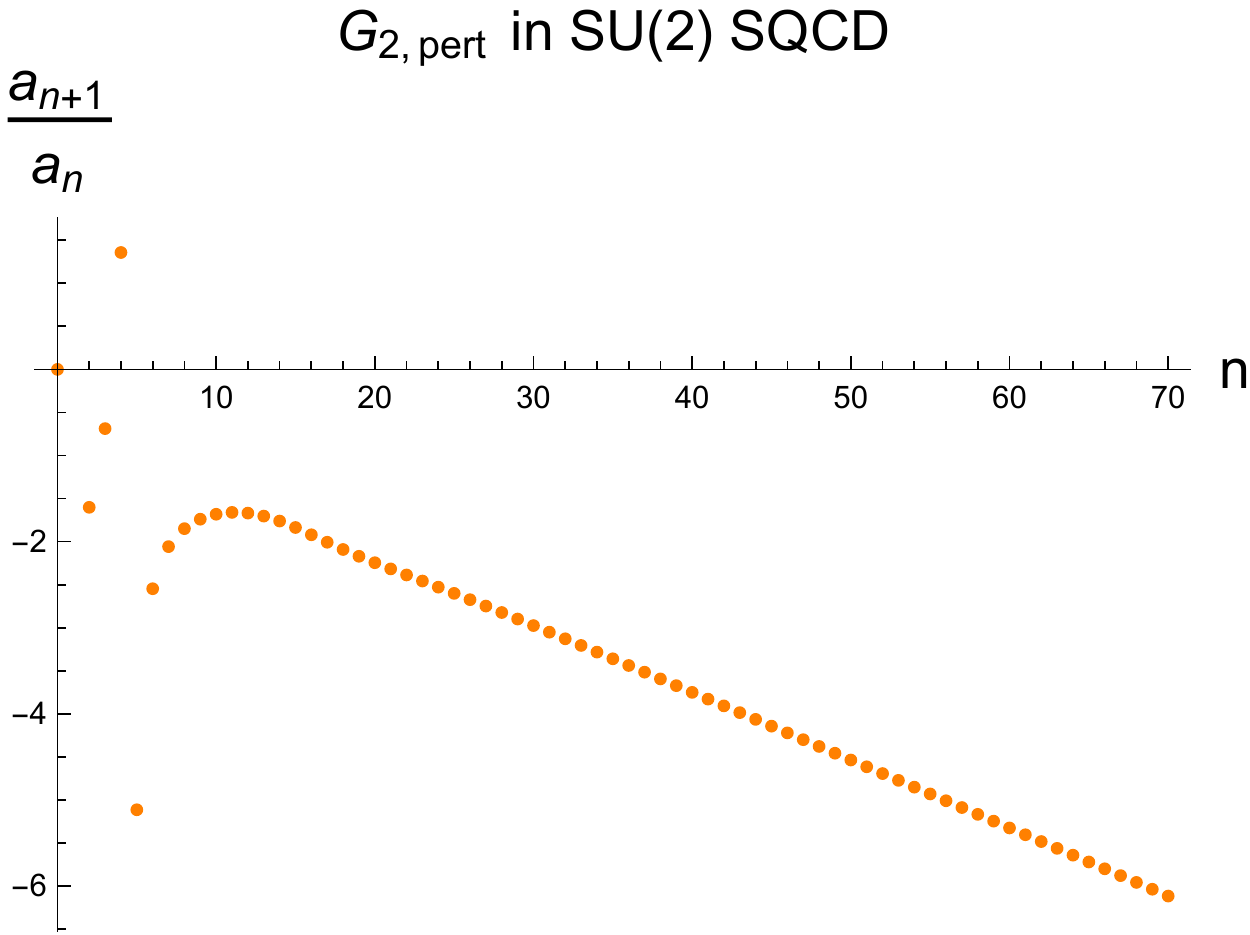}
 \caption{The ratio of consecutive coefficients appearing in the perturbative expansion \eqref{G2Expansion} of $G_2$ in $SU(2)$ SQCD plotted in terms of the loop order $n$.\label{G2SU2}}
 \end{center}
 \end{figure}

\smallskip
 For example, in a perturbative expansion around weak coupling, $G_2$ is
\begin{equation}\label{pert}
G_2(\tau,\bar\tau)_{\text{pert}}=\frac{6}{({\text {Im}}\tau)^2}-\frac{135\zeta(3)}{2\pi^2}\frac{1}{({\text {Im}}\tau)^4}+\frac{1575\zeta(5)}{4\pi^3}\frac{1}{({\text {Im}}\tau)^5}+{\cal O}\left(\frac{1}{({\text {Im}}\tau)^6}\right)\;.
\end{equation} 
The first two terms in this result were checked against an explicit, two-loop computation in~\cite{Baggio:2014ioa}.  If we denote
 \begin{equation}
  G_2(\tau, \bar \tau)_\text{pert} = \frac{6}{({\text {Im}}\tau)^2} \sum_{n=0}^\infty \frac{a_n}{({\text {Im}}\tau)^n} \,, \label{G2Expansion}
 \end{equation}
it is possible to calculate the coefficients $a_n$ up to fairly high order---see Figure~\ref{G2SU2}.  
From this figure it is clear that the ratio $a_{n+1}/a_n$ asymptotically grows linearly with $n$ with a negative coefficient.  In~\cite{Russo:2012kj, Aniceto:2014hoa}   such   behavior was established for the expansion coefficients of the $S^4$ partition function $Z_{S^4}[\tau, \bar \tau]$.  Moreover, it was shown that the perturbative contribution to $Z_{S^4}[\tau, \bar \tau]$ is Borel summable.  Since $G_2$ can be obtained by taking two derivatives of $\ln Z_{S^4}[\tau, \bar \tau]$, it follows that $G_{2, \text{pert}}$ is also Borel summable.  The one-instanton correction to the perturbative result is non-trivial;  it is given by
\begin{equation}
G_2(\tau,\bar\tau)_{\text{1-inst}}=\cos\theta \,e^{-\frac{8\pi^2}{g^2}}\left(\frac{6}{({\text {Im}}\tau)^2}+\frac{3}{\pi}\frac{1}{({\text {Im}}\tau)^3}-\frac{135\zeta(3)}{2\pi^2}\frac{1}{({\text {Im}}\tau)^4}+{\cal O}\left(\frac{1}{({\text {Im}}\tau)^5}\right)\right)\;.
\end{equation}  

\smallskip
The perturbative expression \eqref{G2Expansion} can be used to check the conjecture of \cite{Karliner:1998ge}, originally formulated for the case of QCD\@.  The conjecture is that a Pad\'e approximation of order $(n/2, n/2)$ obtained from the $n$-loop result (with $n$ even) can be used to estimate the value of $a_{n+1}$ with exponentially small error.  If we denote the estimate of $a_{n+1}$ using the symmetric Pad\'e by $a_{n+1, \text{estimated}}$, the conjecture is that
 \begin{equation}
  \left| \frac{a_{n+1, \text{estimated}}}{a_{n+1}}-1 \right| < C e^{-\sigma n} \label{Conv}
 \end{equation}
for some $\sigma > 0$ and $C>0$. 
 \begin{figure}[t]
\begin{center}
\includegraphics[width = 0.5\textwidth]{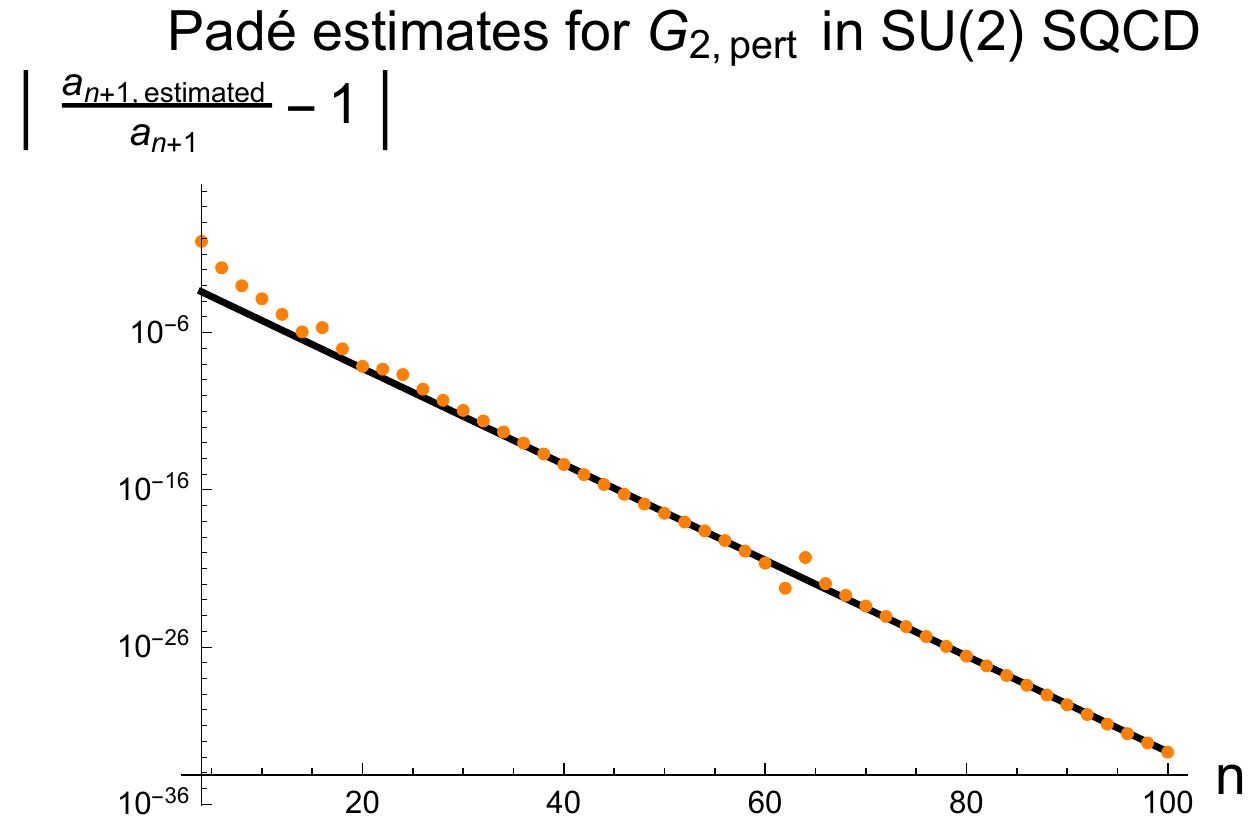}
 \caption{The relative difference between the Pad\'e estimate of the coefficient $a_{n+1}$ and its actual value in the case of $G_2$ in $SU(2)$ SQCD\@. The black line is a linear fit for $n \geq 40$. \label{PADESU2}}
 \end{center}
 \end{figure}
As we show in Figure~\ref{PADESU2}, the relation \eqref{Conv} is indeed true, with an exponent $\sigma \approx 0.7$ that can be determined from the slope of the logarithmic plot.\footnote{If the perturbative series was simply
$a_n=(-1)^nn!$, then~\eqref{Conv} would have been satisfied with $\sigma=\ln(2)$. Even though the situation here is more complicated and there are in fact infinitely many poles on the negative axis of the Borel plane~\cite{Russo:2012kj, Aniceto:2014hoa}, we still seem to find $\sigma\sim \ln(2)$. It would be interesting to understand this better.  }

\subsection{$SU(N)$ Gauge Group}

SCFTs based on a single $SU(N)$ gauge group have one exactly marginal coupling constant, $\tau$, as in the $SU(2)$ case. The chiral ring is generated by the $N-1$ operators
$\phi_k=i^{k+1}(4\pi)^{k/2} \Tr(\varphi^k),\ k=2,\ldots, N$.
The dimension-two operator $\phi_2$, as usual, corresponds to the exactly marginal deformation. 
We can use the following basis in the space of chiral operators
\beq
\cO_{\{n_i\}}=\prod_{k=2}^N\left(\phi_k\right)^{n_k}\;.\label{basisa}
\eeq
\smallskip
 
In order to implement the algorithm of subsection 2.4. in these theories, we  first deform the SCFT action on $S^4$ by 
$$S_{\text{SCFT}}\rightarrow S_{\text{SCFT}}-{1\over 32\pi^2}\int d^4x\, d^4\theta\,{\cal E}\, \sum_{\frak{a}=3}^N \tau^{A}\phi_{A}~, $$
 and compute the $S^4$ partition function $Z[S^4](\tau,\bar \tau; \tau^{A},\bar \tau^{A})$ of the deformed SCFT. 

\smallskip

We are interested in the two-point functions in flat space $\langle \cO_{\{n_i\}}(0)\overline{{\cO}}_{\{n'_i\}}(\infty)\rangle_{\bR^4}$.  These are potentially non-vanishing if
$\sum_{k=2}^N kn_k=\sum_{k=2}^N kn'_k$. Note that unlike in the case  of $SU(2)$, for higher rank gauge group,  there can be more than one operator of a given dimension and hence mixing already on $\bR^4$, for example between $(\Tr(\varphi^3))^2$ and $(\Tr(\varphi^2))^3$.
\smallskip

As before, we begin by studying the matrix of two-point functions on $S^4$ $\cM_{\{n_i\},\{n'_i\}}=\langle \cO_{\{n_i\}}(N)\overline{{\cO}}_{\{n'_i\}}(S)\rangle_{S^4}$. On $S^4$, this matrix  is in general nonzero for all $\{n_i\},\{n'_i\}$. We could compute the correlators from the $S^4$ partition function $Z[S^4](\tau,\bar \tau; \tau^{A},\bar \tau^{A})$
by taking derivatives 
\beq\label{foursp} \cM_{\{n_i\},\{n'_i\}}= {1\over Z[{S^4}]}{\partial^{n_2}\over (\partial \tau)^{n_2}}{\partial^{n_3}\over (\partial \tau^3)^{n_3}}\cdots {\partial^{n_N}\over (\partial \tau^N)^{n_N}}{\partial^{n'_2}\over (\partial \bar\tau)^{n'_2}}{\partial^{n'_3}\over (\partial\bar  \tau^3)^{n'_3}}\cdots {\partial^{n'_N}\over (\partial \bar \tau^N)^{n'_N}}Z[{S^4}](\tau,\bar \tau; \tau^{A},\bar \tau^{A})\biggr|_{\tau^{A}=\bar \tau^{A}=0}~.\eeq
Then, we perform the Gram-Schmidt procedure to extract the two-point functions in flat space $\langle \cO_{\{n_i\}}(0)\overline{{\cO}}_{\{n'_i\}}(\infty)\rangle_{\bR^4}$.

\smallskip

We can explicitly   determine the chiral ring data to all orders in perturbation theory.
Unfortunately, the  expression for $Z_{\Omega,\text{inst}}(a,\tau^i,\tau^A)$ that appears in \rf{deformint} and \rf{productp} in the localization computation of $Z[S^4](\tau,\bar \tau; \tau^{A},\bar \tau^{A})$ in   $SU(N)$  SQCD   with $2N$ fundamental hypermultiplets  is not yet available in the literature. The dependence of the instanton partition function on the higher Casimir couplings, $\tau^{A}$ ($A=3,..,N$) is unknown. (While it is available  for $U(N)$ theories~\cite{Losev:2003py,Nekrasov:2003rj,Nakajima:2003uh,Fucito:2015ofa}  it is an open problem to compute them for $SU(N)$.) Ignoring the instantons, one can  nevertheless use~\eqref{foursp} to derive many interesting results independent of the specific expression for $Z[S^4](\tau,\bar \tau; \tau^{A},\bar \tau^{A})$. One application is the derivation of the coupled $tt^*$ equations which are  obtained from~\eqref{foursp} in Appendix~\ref{couptt}.  The general $tt^*$ equations were  first derived in~\cite{Papadodimas:2009eu}. In addition, we can say quite a bit about the structure of the solution to the $tt^*$ equations in the case of SQCD.

\subsubsection{$SU(N)$  with an Adjoint Hypermultiplet}
This corresponds to the maximally supersymmetric $\mathcal{N}=4$ super-Yang-Mills theory. In this theory, the four-sphere partition receives no instanton corrections~\cite{Okuda:2010ke}. The deformed partition function is  given by a quadratic matrix model deformed by the higher Casimirs evaluated on the localization locus  (our discussion here can be generalized to any gauge group):  
\begin{equation} \label{N4gen} 
Z[S^4](\tau,\bar{\tau}; \tau^{A},\bar \tau^{A})=\int d^{N-1}a\,\Delta(a) |e^{i \pi  \tau \,\Tr(a^2)+i\sum_{A=3}^N \pi^{A/2}\tau^{A} \Tr(a^{A})  }|^2~.
\end{equation}
 
\smallskip 
We  show that the relatively simple form~\eqref{N4gen} leads to two main consequences which we  now derive

\begin{itemize}
\item The flat-space two-point functions $\langle \cO_{\{n_i\}}(0)\overline{{\cO}}_{\{n'_i\}}(\infty)\rangle_{\bR^4}$ are saturated by tree diagrams. This is a trivial consequence of the form of~\eqref{N4gen}. This property of $\mathcal{N}=4$ is further discussed in \cite{Baggio:2012rr,Lee:1998bxa,DHoker:1999ea}.
\item The chiral ring data can be organized in terms of infinitely many decoupled Toda chains.
\end{itemize}

Both of these conclusions are special to $\mathcal{N}=4$ super-Yang-Mills. As we will see below, the second conclusion is actually also valid in other theories up to two loops but not to higher orders.

\smallskip

In order to establish the second point we need to make some simple observations.
The first observation is that multiplying two orthogonal operators  that do not explicitly depend on $\tau$  by powers of $\phi_2$ does not change the fact that they are orthogonal: 
\begin{equation} \vev{\cO_I(N)\overline \cO_J(S)}_{S^4}=0\;\; \Rightarrow \;\;\vev{\phi^n_2\cO_I(N)\overline{\phi_2^m\cO_J}(S)}_{S^4}=0\;.\label{c2-descendents-orthogonal-sphere}
\end{equation} 
This follows from~\eqref{N4gen}. Indeed, if two operators are orthogonal and if they are independent of $\tau$, then by taking derivatives with respect to $\tau,\bar \tau$ one finds~\eqref{c2-descendents-orthogonal-sphere}.

\smallskip 
Thus, if we choose a basis of the form \begin{equation}
\cO^{(m)}_n=\phi_2^n\cO^{(m)}_0\;,
\label{basis}
\end{equation} with the operators  $\cO_0^{(m)}$ constructed such that they are orthogonal to each other
\begin{equation}\vev{\cO_0^{(m)}(N)\overline{\cO_0^{(m')}}(S)}_{S^4}=0\;,\;\;\; \text{for}\;m\neq m'\;,\label{orthogonality_requirement}
\end{equation}
and such that $\cO^{(m)}_0$ do not explicitly depend on $\tau$, then in the basis~\rf{basis}
our system splits into orthogonal sectors:
\begin{equation}\vev{\cO_n^{(m)}(N)\overline{\cO_k^{(m')}}(S)}_{S^4}=0\;,\;\;\; \text{for}\;m\neq m'\;.\label{orthogonality}
\end{equation}
In~\cite{Baggio:2015vxa}, the operators $\cO_0^{(m)}$ are called $C_2$ primaries because they have, in a sense, the minimal possible number of $\phi_2$ factors. 

\smallskip

It is easy to construct the basis~\rf{basis} explicitly and verify that the operators in it are independent of $\tau$. This is done as follows. We consider the set of operators of the form $\prod_{k=3}^N\left(\phi_k\right)^{n_k}$ (i.e. operators from the basis \eqref{basis} with $n_2=0$), and choose an ordering on this set such that the operators are labeled as ${B_m}$, with $\Delta_{m}\leq\Delta_{m+1}$ (thus, $B_0=\mathds{1}$, $B_1=\phi_3$,...). We can now define  $\cO^{(m)}_0$ by an inductive process. For $m=0$,  we choose $\cO^{(0)}_0=B_0=\mathds{1}$. Assuming that we have defined $\cO^{(m')}_0$  with $m'$ ranging from $0$   up to $m-1$, we define $\cO^{(m)}_0$ to be a linear combination of $B_m$ and operators of the form $\cO^{(m')}_{n_{m'}}=\phi_2^{n_{m'}}O^{(m')}_0$, where $m'<m$ and ${n_{m'}}=\frac{\Delta_m-\Delta_m'}{2}$ is an integer. Note that $B_m$ and  $\cO^{(m')}_{n_{m'}}$   have the same dimension $\Delta_m$. This fact will be important to us soon.  The coefficients in this linear combination are chosen such that $\vev{\cO_0^{(m)}(N)\overline{\cO_0^{(m')}}(S)}_{S^4}=0$ will be obeyed for all $m'<m$, that is,  
\begin{equation}\label{mixing4}
\cO^{(m)}_0=B_m-\sum_{m'}\frac{\vev{ B_m(N) \overline \cO^{(m')}_0(S)}_{S^4}}{\left\langle \cO^{(m')}_{n_{m'}}(N) \overline \cO^{(m')}_0(S)\right\rangle}_{S^4} \cO^{(m')}_{n_{m'}}\;,
\end{equation} 
where the sum above is only on $m'$ such that ${n_{m'}}=\frac{\Delta_m-\Delta_m'}{2}\in \mathbb{N}$. This construction makes it obvious that the $\cO^{(m)}_0$ are $\tau$-independent, as required. Indeed, since  the coefficients in~\eqref{mixing4} have the same dimension in the numerator and the denominator, and  since these correlators in $\cN=4$ super-Yang-Mills are tree-level exact, the factors of $\tau$ cancel.    In summary, we have constructed a basis of operators in the chiral ring that decouple into mutually orthogonal semi-infinite towers   whose   bottom operators are explicitly $\tau$-independent.

\smallskip

For example, the first towers in $SU(N)$ for $N\geq4$ are  
\begin{equation}
O^{(0)}_n=\phi_2^n\;,\;\;\; O^{(1)}_n=\phi_2^n\phi_3\;,\;\;\; O^{(2)}_n=\phi_2^n\left(\phi_4-\frac{\vev{\phi_4(N)}_{S^4}}{\vev{\phi_2^2(N)}_{S^4}}\phi_2^2\right)\;\ldots .
\end{equation}   
By construction, this new basis satisfies (\ref{orthogonality})  and as a result one can perform the Gram-Schmidt procedure of subsection \ref{summary} in each tower separately. This leads to  a tremendous simplification.
If we denote the matrix elements in this basis by $M^{(m)}_{i,j}=\vev{O^{(m)}_i(N)\overline{O^{(m)}_j}(S)}_{S^4}$, exactly the same derivation as the one presented in the case of SCFTs based on $SU(2)$ proves that the chiral data, encoded in $G^{(m)}_{2n}\equiv \vev{O^{(m)}_n(0)\overline{O^{(m)}_n}(\infty)}_{\bR^4}$,  satisfies
\begin{equation}
\begin{aligned}\label{toda4}
&16\,\partial_\tau\partial_{\bar{\tau}}\ln G^{(m)}_{2n}=\frac{G^{(m)}_{2n+2}}{G^{(m)}_{2n}}-\frac{G^{(m)}_{2n}}{G^{(m)}_{2n-2}}-G_2\;,\\
&16\,\partial_\tau\partial_{\bar{\tau}}\ln G^{(m)}_{0}=\frac{G^{(m)}_{2}}{G^{(m)}_{0}}-G_2\;,
\end{aligned}
\end{equation}
and  $G_2=16\,\partial_\tau \partial_{\bar\tau} \ln Z[S^4](\tau,\bar{\tau}; 0,0)$.

\smallskip

Equation~\eqref{toda4} describes decoupled semi-infinite Toda chains, in agreement with~\cite{Baggio:2015vxa}.
One can explicitly solve for the $G^{(m)}_{2n}$ using the fact that 
\begin{equation}
G_2=\frac{2(N^2-1)}{({\text {Im}}\tau)^2}\;.
\end{equation}
One finds 
\begin{equation}
G^{(m)}_{2n}(\tau,\bar{\tau})=4^{n}\frac{n!\,\tilde G^{(m)}_0}{({\text {Im}}\tau)^{\Delta_m+2n}}\left(\frac{N^2-1}{2}+\Delta_m\right)_n\;,
\end{equation}
where $(x)_n$ is  the Pochhammer symbol 
\eq{(x)_n=x(x+1)...(x+n-1)\ }
and $\tilde G^{(m)}_0$ encodes the normalization of the operator at the bottom of the $m$-th tower.

\smallskip
As we have already emphasized, this structure of decoupled Toda chains obviously exists at tree-level in $\mathcal{N}=2$ $SU(N)$ SQCD as well (actually, in any SCFT at tree level). As we will show in the next subsection, it persists up to two-loops in $SU(N)$ SQCD.

\subsubsection{Decoupled Toda Chains at Two-Loops in SQCD}

We now show  that  the decoupled Toda chain structure~\eqref{toda4} remains in $SU(N)$ SQCD   up to two-loops.  That is, the chiral ring data can be organized in terms of decoupled semi-infinite Toda chains up to that order in perturbation theory.

\smallskip
The operators $\cO_0^{(m)}$ constructed in~\eqref{mixing4} are orthogonal at tree-level, but they are not guaranteed to stay orthogonal when higher-order corrections are included. If the operators were to stay orthogonal for all values of the coupling constant,  then equation (\ref{toda4}) would hold to all orders. Let us explain why the first non-trivial two-loop correction actually does not ruin the orthogonality that was achieved at tree-level. 

\smallskip

The first non-trivial perturbative correction can be obtained by expanding the  
  matrix integral representation (see section \rf{sec:defpart})\footnote{The perturbative matrix integral of $SU(N)$ SQCD is
  	\begin{equation}  
  	Z[S^4](\tau,\bar{\tau},...)=\int d^{N-1}a\, e^{-2\pi\, {\text{Im}}\tau \,\Tr a^2+...} \prod_{i\neq j}( w_{ij}\cdot a)\, {\prod_{i\neq j}H(i w_{ij}\cdot a) \prod_i H(iw_i\cdot a)^{-2N}}~,
  	\end{equation}
  	where $w_i$, $i=1,..,N$, are the weights in the fundamental representation, and $w_{ij}=w_i-w_j$.
  } of the deformed SCFT partition function on $S^4$ 
\begin{equation}
\begin{aligned}
&\frac{1}{Z[S^4]}\int d^{N-1}a\,\Delta({a})F( a)e^{-2\pi\, {\text{Im}}\tau\,\Tr{a}^2}\left(1-3\zeta(3)(\Tr a^2)^2\right)=
\\	&\frac{1}{Z[S^4]}\left(1-3\zeta(3)\frac{\partial^2}{\partial (2\pi\, {\text{Im}}\tau)^2}\right)\int d^{N-1}{a}\,\Delta({a})F( a)e^{-2\pi\, {\text{Im}}\tau\,\Tr{a}^2}\ ,
\end{aligned}
\end{equation}
where $F(a)$ denotes some  insertion in the localization formula. 

\smallskip
The fact that the first non-trivial correction is obtained from the tree-level result by differentiating with respect to the coupling constant of the theory implies that the towers constructed to be orthogonal at tree-level~\eqref{orthogonality}  will remain orthogonal also  up to two-loops  in perturbation theory. 

\smallskip

There is no reason to expect that this property will be true also for the next orders, and indeed the results that we present next contradict the decoupling conjecture already at the next order in perturbation theory.

\subsection{$SU(3)$ and $SU(4)$ SQCD}
\subsubsection{$SU(3)$ SQCD}
We consider $SU(3)$ SQCD  with 6 fundamental hypermultiplets to  three-loops. 
We show that at this   order in perturbation theory   
the  bottom operators of the towers we constructed before become  explicitly $\tau$-dependent.
This indicates that there is no reason to expect decoupled Toda chains as in~\eqref{toda4} anymore. 

\smallskip

Let us consider the first few low-lying chiral operators in $SU(3)$ SQCD: $\{\phi_2\}$, $\{\phi_3\}$, $\{{\phi_2}^2\}$, $\{{\phi_3\phi_2}\}$, $\{{\phi_2}^3,{\phi_3}^2\}$, $\{\phi_3{\phi_2}^2\}$ and $\{{\phi_2}^4,{\phi_3}^2\phi_2\}$. We are interested in their two-point functions in flat space
$(G_{\Delta})_{IJ}=\vev{O_{\Delta I}(0)\overline{O_{\Delta J}}(\infty)}_{\bR^4}$.  $G_6$ and $G_8$ are therefore $2\times 2$ matrices while the rest are just functions of the gauge coupling $g$ in perturbation theory. Following our Gram-Schmidt procedure we can compute these  up to three-loops 
\begin{align}
G_2 & = \left(\frac{g^2}{4\pi}\right)^2\left(16 - \frac{45 \,\zeta(3)}{2\pi^4} g^4+\frac{ 425\,\zeta(5)}{8\pi^6} g^6 + O(g^8)\right)~,\\[5px]
G_3 & = \left(\frac{g^2}{4\pi}\right)^3\left(40 - \frac{135 \,\zeta(3)}{2\pi^4} g^4+\frac{6275\,\zeta(5)}{48\pi^6}g^6 + O(g^8)\right)~,\\[5px]
G_4 & = \left(\frac{g^2}{4\pi}\right)^4\left(640 - \frac{2160 \,\zeta(3)}{\pi^4} g^4+\frac{6375 \,\zeta(5)}{\pi^6} g^6+ O(g^8)\right)~,\\[5px]
G_5 & = \left(\frac{g^2}{4\pi}\right)^5\left(1120 - \frac{4410 \,\zeta(3)}{\pi^4} g^4+\frac{144725\,\zeta(5)}{12\pi^{6}}g^6+ O(g^8)\right)~,\\[5px]
G_6 & = \left(\frac{g^2}{4\pi}\right)^6
\begin{pmatrix}
46080 - \frac{272160 \,\zeta(3)}{\pi^4} g^4 +\frac{969000 \,\zeta(5)}{\pi^6} g^6&~ i\left(1920 - \frac{11340 \,\zeta(3)}{\pi^4} g^4 +\frac{29875 \,\zeta(5)}{\pi^6} g^6\right)\\[5px]
-i\left(1920 - \frac{11340 \,\zeta(3)}{\pi^4} g^4 +\frac{29875 \,\zeta(5)}{\pi^6} g^6\right) &~ 6800 - \frac{57645 \,\zeta(3)}{2\pi^4} g^4+\frac{1688875 \,\zeta(5)}{24\pi^6} g^6
\end{pmatrix} \notag \\
&+O(g^{20}) \,,
\end{align}
\begin{align}
G_7 & = \left(\frac{g^2}{4\pi}\right)^7\left(71680 - \frac{483840 \,\zeta(3)}{\pi^4} g^4+\frac{4936400\,\zeta(5)}{3\pi^6}g^6+ O(g^8)\right)~,\\[5px]
G_8 &  = \left(\frac{g^2}{4\pi}\right)^8
\begin{pmatrix}
5160960 - \frac{46448640 \,\zeta(3)}{\pi^4}g^4+\frac{194208000 \,\zeta(5)}{\pi^6} g^6 &~ i\left(215040 - \frac{1935360 \,\zeta(3)}{\pi^4} g^4+\frac{6412000 \,\zeta(5)}{\pi^6} g^6\right) \\[5px]
-i\left(215040 - \frac{1935360 \,\zeta(3)}{\pi^4} g^4+\frac{6412000 \,\zeta(5)}{\pi^6} g^6 \right)&~ 277760 - \frac{2046240 \,\zeta(3)}{\pi^4}g^4+\frac{20027000 \,\zeta(5)}{3\pi^6} g^6 
\end{pmatrix} \notag\\
&+O(g^{24}) \,.
\end{align}
This is in agreement with \cite{Baggio:2015vxa}, where the same correlators were computed up to two-loops using standard perturbation theory. 
It would be interesting to verify our three-loop results by direct perturbative computations.
\smallskip

We now note that in performing the Gram-Schmidt procedure on the dimension $6$ and $8$ operators, we encounter the following ratios
\begin{equation}
\frac{G_6(2,1)}{G_6(1,1)}=-i\left(\onov{24}-\frac{175\,\zeta(5)}{768\pi^6}\,g^6+\CO(g^{8})\right)\ ,\label{hhh}
\end{equation}
\begin{equation}
\frac{G_8(2,1)}{G_8(1,1)}=-i\left(\onov{24}-\frac{125\,\zeta(5)}{384\pi^6}\,g^6+\CO(g^{8})\right)\ .\label{jjj}
\end{equation}
The presence of the $g^6$ term means that one cannot diagonalize $G_4$ and $G_6$ with a  $\tau,\bar{\tau}$ independent  basis. This contradicts the conjecture of~\cite{Baggio:2015vxa}. 
Note that the absence of a term $g^4$ in~\eqref{hhh},\eqref{jjj} is precisely as anticipated in 3.2.2. The decoupling of Toda chains therefore starts to fail  at three-loop order in perturbation theory.
\smallskip

\begin{figure}[t]
\begin{center}
\includegraphics[width = 0.49\textwidth]{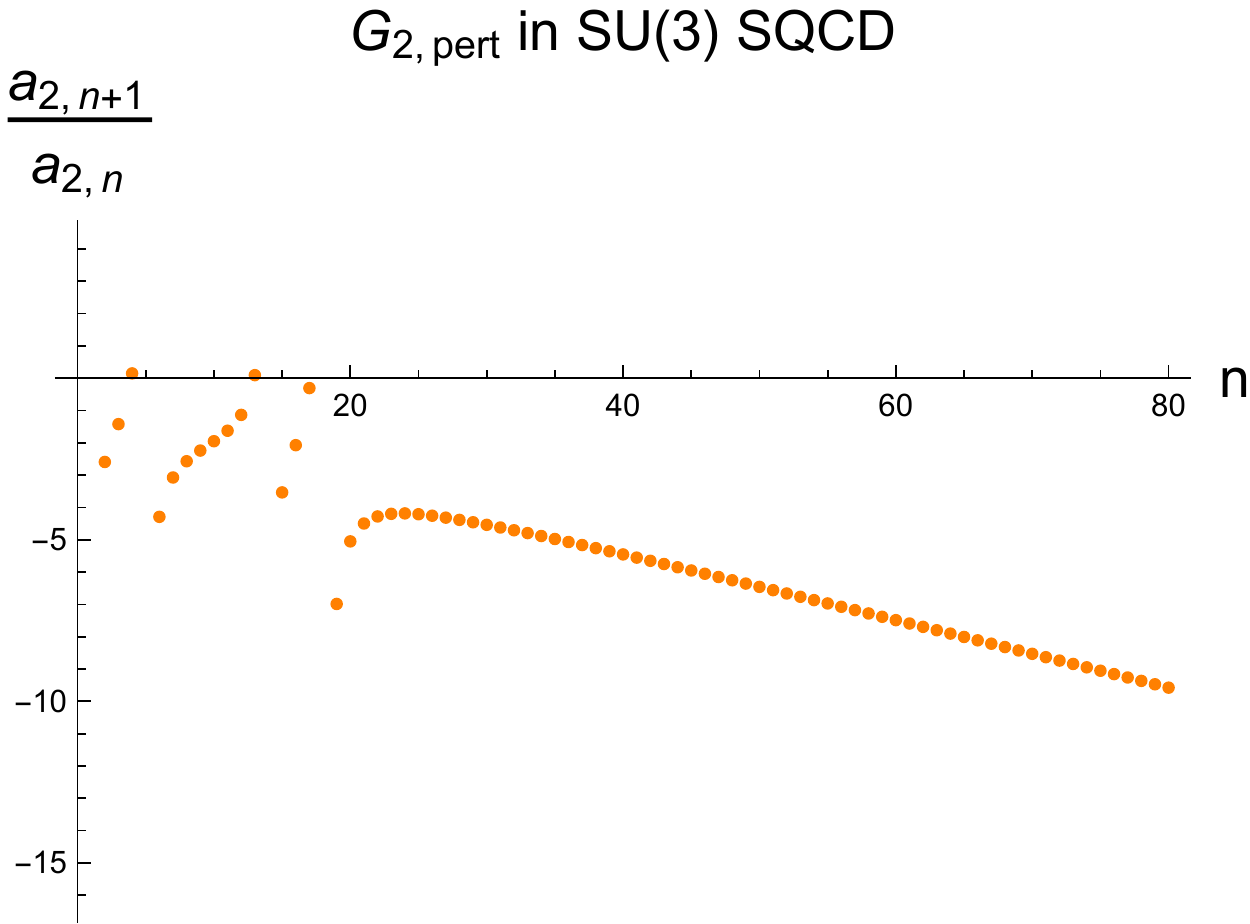}
\includegraphics[width = 0.49\textwidth]{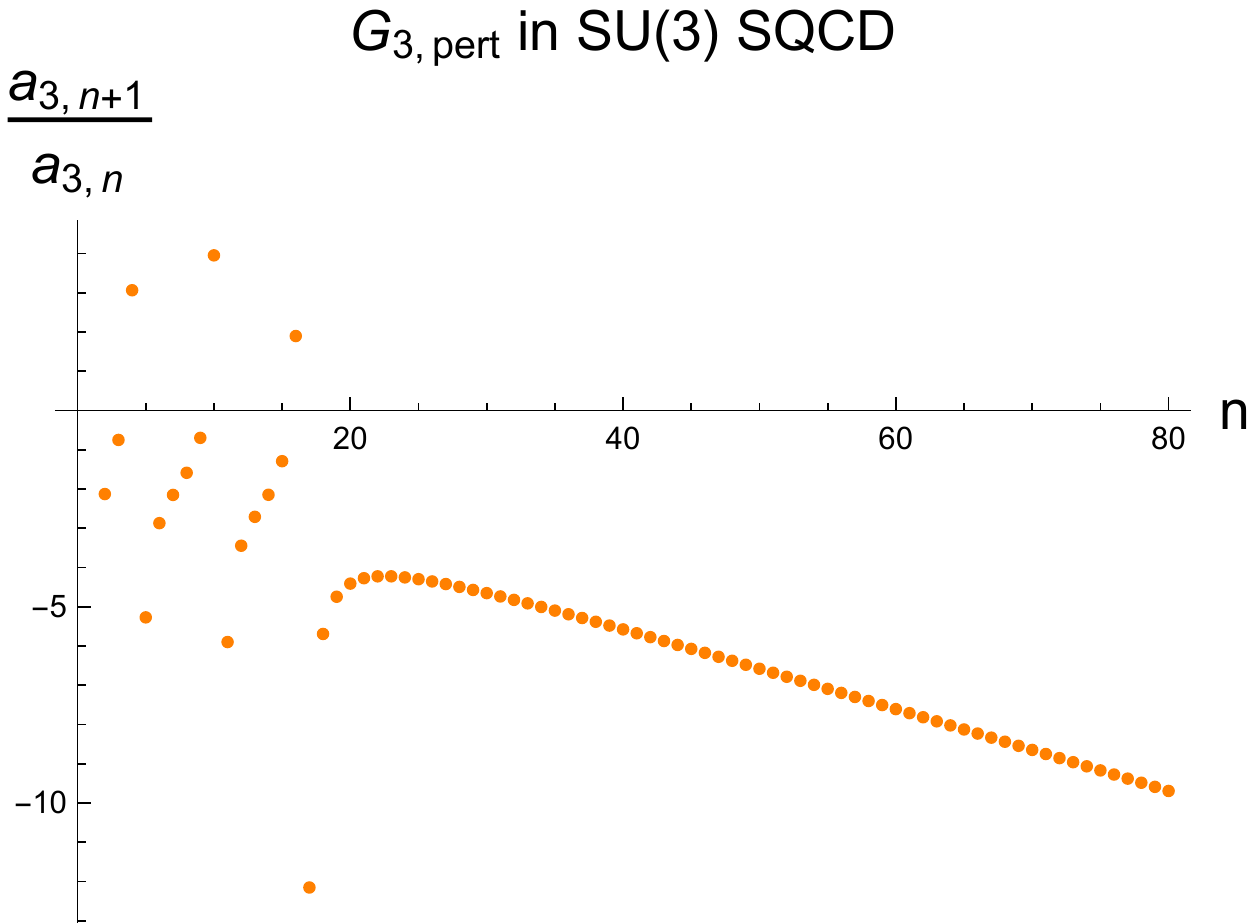}
 \caption{Ratios of consecutive coefficients in the series expansions \eqref{GnExpansion} in the case of $SU(3)$ SQCD\@.\label{RATIOSU3}}
 \end{center}
 \end{figure}

  If we define
 \begin{equation}
  G_{m,\text{pert}}(g^2) = G_{m, \text{tree}} \sum_{n=0}^\infty a_{m, n} \left( \frac{g^2}{4 \pi} \right)^n \label{GnExpansion}
 \end{equation}
 where $G_{m, \text{tree}}$ is the tree-level contribution and so $a_0 = 1$, one can check that the ratio $a_{m, n+1} / a_{m, n}$ grows linearly at large $n$ with a negative coefficient, just as was the case for $SU(2)$ SQCD\@.  See Figure~\ref{RATIOSU3} for plots of these ratios in the cases $m=2, 3$.  We expect that $G_{m, \text{pert}}$ is also Borel summable in this case, but we have not shown this conclusively.   
\smallskip

\begin{figure}[t]
\begin{center}
\includegraphics[width = 0.49\textwidth]{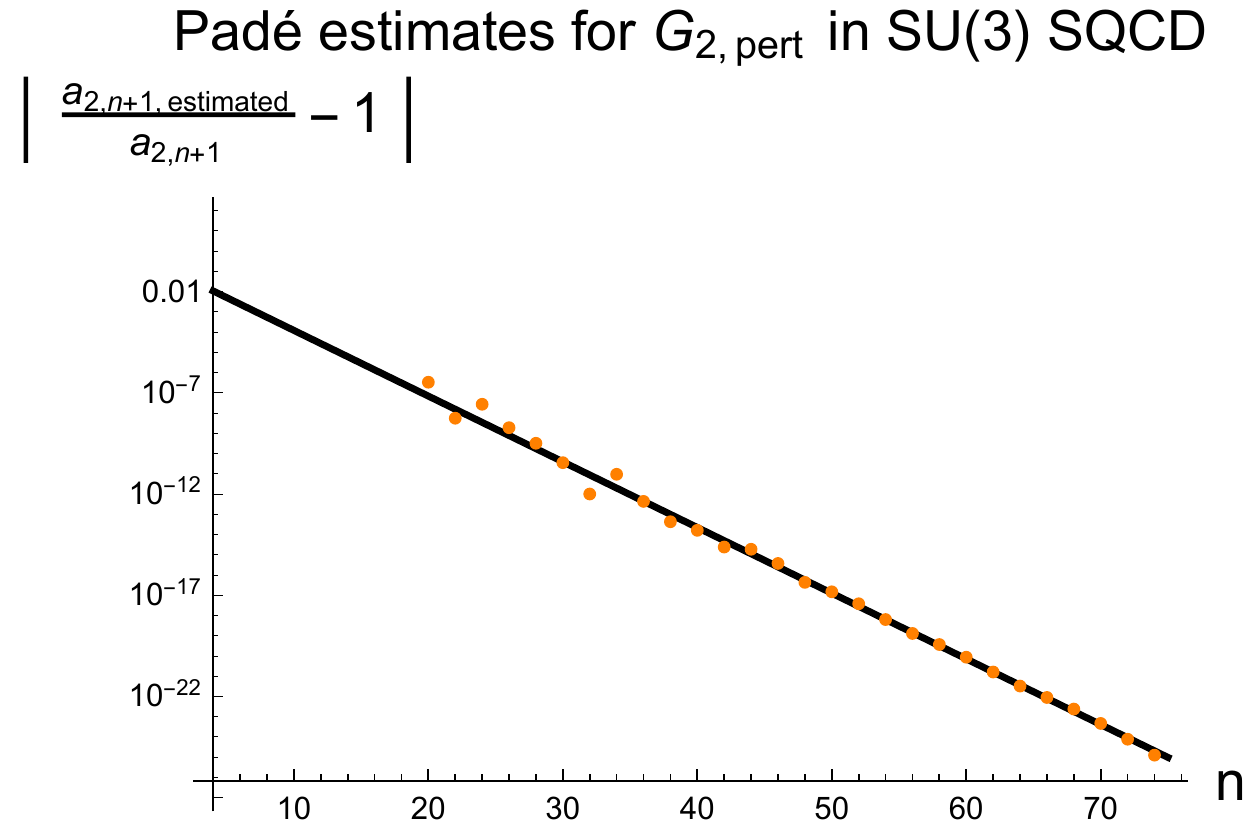}
\includegraphics[width = 0.49\textwidth]{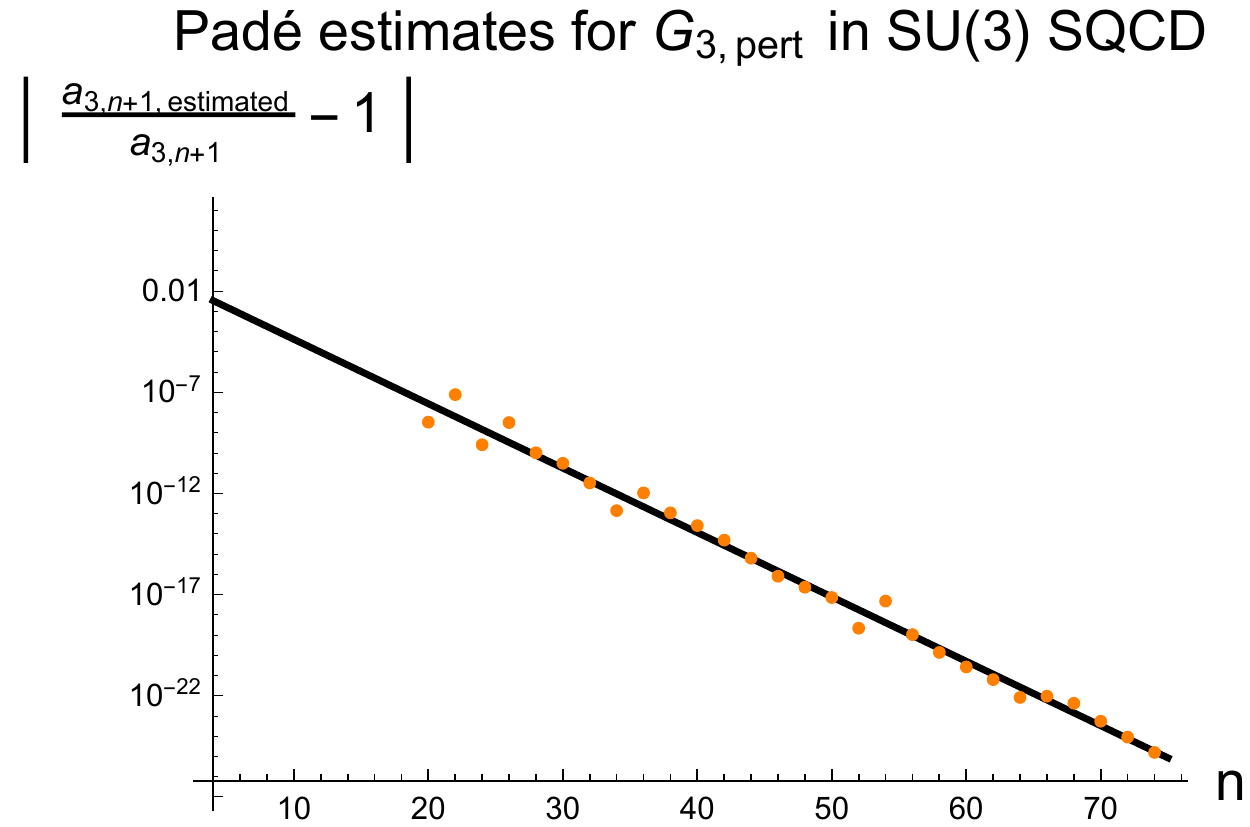}
 \caption{The relative difference between the Pad\'e estimate of the coefficient $a_{m, n+1}$ and its actual value in the case of $G_2$ and $G_3$ in $SU(3)$ SQCD\@. The black lines are linear fits for $n \geq 40$. \label{PADESU3}}
 \end{center}
 \end{figure}

\smallskip
As in the case of $SU(2)$ SQCD, one can use the series expansions above to estimate whether the $(n/2, n/2)$ Pad\'e, computed only from the first $n$ terms, can be used to estimate the $(n+1)$th series coefficient with an exponentially small error.  This is indeed the case, as can be seen from Figure~\ref{PADESU3} for $m=2, 3$.  
Defining the exponents $\sigma_m$ through
 \begin{equation}
  \left| \frac{a_{m, n+1, \text{estimated}}}{a_{m, n+1}}-1 \right| < C_m e^{-\sigma_m n} \label{ConvGen}
 \end{equation}
linear fits of the log plots in Figure~\ref{PADESU3} give $\sigma_2 \approx 0.75$ and $\sigma_3 \approx 0.73$.  These values are rather close to the corresponding values for $SU(2)$ SQCD\@.

\subsubsection{$SU(4)$ SQCD}
The conclusion from our study of $SU(4)$ SQCD is the same as the conclusion from the study of $SU(3)$ SQCD above. We present it just in order to demonstrate again the cancelation of the $g^4$ term and to provide additional data that can be compared with direct perturbative computations. 
We consider the operators  $\{\phi_2\}$, $\{{\phi_3}\}$,
$\{{\phi_2}^2,{\phi_4}\}$ and $\{{\phi_2}{\phi_3}\}$ and denote the corresponding two-point functions by $G_2$, $G_3$, $G_4$ and $G_5$, respectively. Using our Gram-Schmidt procedure we find

\begin{align}
G_2 & = \left(\frac{g^2}{4\pi}\right)^2\left(30 - \frac{2295 \,\zeta(3)}{32\pi^4} g^4+\frac{118575\,\zeta(5)}{512\pi^6}g^6 + O(g^8)\right)~,\\[5px]
G_3 & = \left(\frac{g^2}{4\pi}\right)^3\left(135 - \frac{23085 \,\zeta(3)}{64\pi^4} g^4+\frac{4100625\,\zeta(5)}{4096\pi^6}g^6 + O(g^8)\right)~,\\[5px]
G_4 & =\left(\frac{g^2}{4\pi}\right)^4
\begin{pmatrix}
2040 - \frac{43605 \,\zeta(3)}{4\pi^4} g^4 +\frac{1304325\,\zeta(5)}{32\pi^6}g^6 &~ i\left(870 - \frac{74385 \,\zeta(3)}{16\pi^4} g^4 +\frac{2351025\,\zeta(5)}{128\pi^6}g^6\right)\\[5px]
-i\left(870 - \frac{74385 \,\zeta(3)}{16\pi^4} g^4 +\frac{2351025\,\zeta(5)}{128\pi^6}g^6 \right) &~ \frac{1335}{2} - \frac{198045 \,\zeta(3)}{64\pi^4} g^4+\frac{5681925\,\zeta(5)}{512\pi^6}g^6 
\end{pmatrix} \notag \\
 &+ O(g^{16})\\[5px]
G_5 & = \left(\frac{g^2}{4\pi}\right)^5\left(5670 - \frac{535815 \,\zeta(3)}{16\pi^4} g^4+\frac{248558625\,\zeta(5)}{2048\pi^6}g^6 + O(g^8)\right)~.
\end{align}
Again, the two-loop results agree with those that were found by a direct Feynman diagrams computation in  \cite{Baggio:2015vxa}.
In performing  the Gram-Schmidt procedure on the dimension $4$ operators, we encounter the following ratios
\begin{equation}
\frac{G_4(2,1)}{G_4(1,1)}=-i\left(\frac{29}{68}+\frac{525\,\zeta(5)}{1088\pi^6}g^6+\CO(g^8)\right)\;.
\end{equation}
As before, the $g^4$ piece cancels as anticipated but the $g^6$ piece contradicts the conjecture of~\cite{Baggio:2015vxa}. Therefore, we do not expect decoupled semi-infinite Toda chains.

\medskip 
\section*{Acknowledgments}

We would like to thank D.~Butter, O.~Foda, D.~Gaiotto,  M.~Karliner, H.~C.~Kim, S.~Komatsu, N.~Nekrasov, V.~Niarchos, K.~Papadodimas, V.~Pestun, L.~Rastelli, S.~Razamat, A.~Schwimmer, N.~Seiberg, and R.~Yacoby for fruitful discussions. This research was supported in part by
Perimeter Institute for Theoretical Physics. Research at Perimeter Institute is supported by the
Government of Canada through Industry Canada and by the Province of Ontario through the Ministry
of Research and Innovation. J.G. also acknowledges further support from an NSERC Discovery
Grant. Z.K. is supported by the ERC STG grant 335182, by the Israel Science Foundation under grant 884/11, by the United States-Israel Bi-national Science Foundation (BSF) under grant 2010/629
as well as by the Israel Science Foundation center for excellence grant (grant no.~1989/14).    Z.K. is also supported by the I-CORE Program of the Planning and Budgeting Committee.  S.S.P. is supported in part by the US NSF under Grant No.~PHY-1418069.  
Any opinions, findings, and conclusions or recommendations expressed in this material are those of the authors and do not necessarily reflect the views of the funding agencies.

\appendix

\section{Integrability of $tt^*$ Equations}  
\label{ap:int}

	In this appendix we will show that  the $tt^*$ equations together with the WDVV equations for  any 4d $\cN=2$ SCFT  are integrable, in the sense that these equations can be written as the flatness condition of a one parameter family of connections on a certain vector bundle over the conformal manifold $\cM$, which is equivalent to the Lax representation (with spectral parameter) of a classically  integrable system. Recall that the WDVV equations~\cite{Witten:1989ig,Dijkgraaf:1990dj,Dijkgraaf:1990qw} and the $tt^*$ equation are~\cite{Papadodimas:2009eu}
	\begin{subequations} \label{WDVVtt*}
	\begin{align}
		& \nabla_i C_{jK}^L = \nabla_j C_{iK}^L, \qquad \overline\nabla_{\bar i} \overline C_{\bar j \bar K}^{\bar L} = \overline\nabla_{\bar j} \overline C_{\bar i \bar K}^{\bar L}, \\
		& [\nabla_i, \nabla_j]_K^L = [\overline\nabla_{\bar i}, \overline\nabla_{\bar j}]_{\bar K}^{\bar L} = 0, \\
		& [\nabla_i, \overline\nabla_j]_K^L = -[C_i, \overline C_j]_K^L + g_{i\bar j} \delta_K^L\left(1 + \frac{R}{4c}\right), \label{tt*}
	\end{align}
	\end{subequations}
%\noindent 
where $i,j$ run over chiral primaries of $\Delta=2$, $K,L$ run over all chiral primaries, and $C_{IJ}^K$ are OPE coefficients defined as:
	\beq
		\cO_I(x) \cO_J(0) = C_{IJ}^K \cO_K(0) + \cdots.
	\eeq
	$C_I$ can be thought of as an operator acting on chiral primaries  whose matrix components are $C_{IJ}^K$. In (\ref{tt*}), $R$ is the R-charge of the chiral primaries that comprise the fibre of the bundle $\mathcal V_R \to \cM$ on which the covariant derivatives and the $C_i$'s act, $c$ is the central charge of the SCFT, and $g_{i\bar j}$ is the Zamolodchikov metric on $\cM$. We also note the fact that $C_I$ ($\overline C_{\bar I}$) is covariantly holomorphic (antiholomorphic)~\cite{Papadodimas:2009eu}
	\beq
		 \overline\nabla_{\bar i} C_I = \nabla_i \overline C_{\bar I} = 0 \label{holo}\,.
	\eeq
	
	\smallskip
	Now consider the holomorphic vector bundle $\mathcal V_R \otimes \mathcal L^{\otimes n}$ with $n = -(4c + R)$, where $\mathcal L$ is the supercharge bundle.\footnote{Negative power of a line bundle is defined as the positive power of the dual bundle, i.e., if $E \to M$ is a line bundle then for some negative real number $m<0$ we have $E^{\otimes m} \equiv {(E^*)}^{\otimes |m|}$.} $\mathcal L$ is a holomorphic line bundle over $\cM$ whose curvature is given by~\cite{Papadodimas:2009eu}
	\beq
		F_{ij} = F_{\bar i \bar j} = 0, \qquad F_{i \bar j} = \frac{1}{4c} g_{i\bar j}. \label{FL}
	\eeq
	This nontrivial curvature encodes the ambiguity in defining the phase of the supercharges, i.e., the following automorphism of the $\cN=2$ superconformal algebra:
	\beq
		Q_\alpha^i \to e^{i\theta} Q_\alpha^i, \quad \overline Q_{\dot\alpha}^{\bar i} \to e^{-i\theta} \overline Q_{\dot\alpha}^{\bar i}, \quad S_\alpha^i \to e^{-i\theta} S_\alpha^i, \quad \overline S_{\dot\alpha}^{\bar i} \to e^{i\theta} \overline S_{\dot\alpha}^{\bar i}.
	\eeq
	(\ref{FL}) implies that the curvature of $\cL^{\otimes n}$, let it be denoted by $F^n$, is given by:
	\beq
		F_{ij}^n = F_{\bar i \bar j}^n = 0, \qquad F_{i \bar j}^n = \frac{n}{4c} g_{i\bar j} = -\left(1+\frac{R}{4c}\right) g_{i\bar j}. \label{FLn}
	\eeq
	Let the covariant derivative on $\cL^{\otimes n}$ be denoted by $\nabla_i^\cL$ and define the following one parameter family of connections on $\mathcal V_R \otimes \cL^{\otimes n}$:
	\beq
		\nabla_i^\xi \equiv \nabla_i + \xi C_i + \nabla_i^\cL, \qquad \overline\nabla_{\bar i}^\xi \equiv \overline\nabla_{\bar i} + \xi^{-1} \overline C_{\bar i} + \overline\nabla_{\bar i}^\cL, \label{conn}
	\eeq
	where $\nabla_i$ and $C_i$ are the same operators that appear in (\ref{WDVVtt*}). The flatness condition of this connection for any value of the parameter $\xi$ is:
	\beq
		[\nabla_i^\xi, \nabla_j^\xi] = [\overline\nabla_{\bar i}^\xi, \overline\nabla_{\bar j}^\xi] = [\nabla_i^\xi, \overline\nabla_{\bar j}^\xi] = 0, \qquad \forall\;\xi\in\mathbb C. \label{flatcon}
	\eeq
	Using (\ref{conn}), (\ref{FLn}) and noting that operators on $\mathcal V_R$ commute with operators on $\cL^{\otimes n}$ we get:
	\begin{subequations} \label{flatx}
	\begin{align}
		[\nabla_i^\xi, \nabla_j^\xi] =&\; [\nabla_i, \nabla_j] + \xi (\nabla_i C_j - \nabla_j C_i) + \xi^2 [C_i, C_j] \\
		[\overline\nabla_{\bar i}^\xi, \overline\nabla_{\bar j}^\xi] =&\; [\overline\nabla_{\bar i}, \overline\nabla_{\bar j}] + \xi^{-1} (\overline\nabla_{\bar i} \overline C_{\bar j} - \overline\nabla_{\bar j} \overline C_{\bar i}) + \xi^{-2} [\overline C_{\bar i}, \overline C_{\bar j}] \\
		[\nabla_i^\xi, \overline\nabla_{\bar j}^\xi] =&\; [\nabla_i, \overline\nabla_{\bar j}] + [C_i, \overline C_{\bar j}] - \left(1+\frac{R}{4c}\right) g_{i\bar j} + \xi^{-1} [\nabla_i, \overline C_{\bar j}] - \xi [\overline\nabla_{\bar j}, C_i] \,.\label{flat3}
	\end{align}
	\end{subequations}
	Equation (\ref{flatcon}) must be satisfied at each order in $\xi$. The $C_i$'s commute among themselves
 and so do the $\overline C_{\bar i}$'s~\cite{Papadodimas:2009eu}, so the $\cO(\xi^2)$ and $\cO(\xi^{-2})$ terms vanish. The $\cO(\xi)$ and $\cO(\xi^{-1})$ terms of (\ref{flat3}) vanish due to (\ref{holo}). By imposing (\ref{flatcon}) order by order on the rest of terms in (\ref{flatx}) we recover precisely (\ref{WDVVtt*}), thus proving integrability of WDVV and $tt^*$ equations of four-dimensional $\cN=2$ SCFTs.  They are  governed by a Hitchin integrable system.

\section{Deforming $\cN=2$ SCFT on $S^4$ by Chiral Operators}  
\label{ap:deform}
	When we place an $\cN=2$ SCFT on $S^4$ via the stereographic map, then the Lagrangian preserves the full superconformal symmetry. However, the partition function and various other observables need to be regulated in the ultraviolet. The maximal subalgebra that can be preserved by the regulator is $osp(2|4)$. This is because this subgroup does not include conformal transformations but only isometries of the sphere. In this appendix we discuss   $F$-term deformations of the action that preserve $osp(2|4)$, i.e.
		\beq
		S \to S - \tau_{{\cal U}} \int_{S^4} \mathrm{d}^4 x \sqrt{g}\, {\cal U}(x)\,,\label{deformation}
	\eeq
	such that the deformation term is $osp(2|4)$ invariant, i.e.
	\beq
		\delta\left(\tau_{\cal U} \int_{S^4} \mathrm{d}^4 x \sqrt{g}\, {\cal U}(x)\right) = 0 \,, \label{symmetric}
	\eeq
	where $\delta$ represents an $osp(2|4)$ transformation. Here $\tau_{\cal U}$ is the coupling constant corresponding to the operator ${\cal U}$. If $\tau_{\cal U}$ has Weyl weight 0 then such a deformation is marginal but here we are interested in more general deformations where $\tau_{\cal U}$ can have arbitrary Weyl weight. The systematic way to find such deformations is to start with an $\cN=2$ superfield and integrate it over the chiral superspace ($\int \mathrm{d}^4 \theta$) with the appropriate measure and evaluate the resulting term on the $S^4$ background. 
	\smallskip
		
	In order to achieve this,  we begin by promoting the coupling constant  $\tau_{\cal U}$ to an $\cN=2$ chiral multiplet with Weyl weight $(2-w)$ whose bottom component is  $\tau_{\cal U}$. We also consider another chiral multiplet of Weyl weight $w$ whose bottom component will be called ${\cal U}$. We will denote these two multiplets as $\underline \tau_{\cal U}$ and $\underline {\cal U}$ and denote the component fields of these two multiplets as $(\tau_{\cal U}, \psi_i, b_{ij}, f_{ab}^-, \lambda_i, c)$ and $({\cal U}, \Psi_i, B_{ij}, F_{ab}^-, \Lambda_i, C)$, respectively.\footnote{$i$, $j$ are $su(2)_R$ indices and $a$, $b$ are local frame indices on $S^4$.} $b_{ij}$, $B_{ij}$ are symmetric and in Euclidean signature $f_{ab}^-$, $F_{ab}^-$ are selfdual tensors. Now the following term is manifestly $osp(2|4)$ invariant:
	\beq
		\int_{S^4} \mathrm{d}^4 x  \int \mathrm{d}^4 \theta\, \cE\, \underline\tau_{\cal U}(x,\theta) \underline{\cal U}(x,\theta) \,, \label{osp24inv}
	\eeq
	where $\cE$ is the chiral density. Since we want $\underline\tau_{\cal U}$ to be a background multiplet, we need to restrict its components in such a way that the required supersymmetry  algebra is unbroken. 
First, in order to preserve rotational invariance on $S^4$ we can give the spacetime scalars $\tau_{\cal U}$, $b_{ij}$ and $c$ constant expectation values and let all the other fields in $\underline \tau_{\cal U}$ vanish. Supersymmetry is preserved if the supersymmetry variations of all the background fields vanish. 

\smallskip

The SUSY variations of a chiral multiplet of Weyl weight $w$, with component fields written as $(A, \Psi_i, B_{ij}, F_{ab}^-, \Lambda_i, C)$, under an $\cN=2$ superconformal transformation are given by~(see e.g. \cite{Gerchkovitz:2014gta}):
	\begin{subequations}\begin{align}
		\delta  A &= \frac{1}{2} \overline\epsilon^i \Psi_i \\
		\delta \Psi_i &= \D(A\epsilon_i) + \frac{1}{2} B_{ij} \epsilon^j + \frac{1}{4} \gamma^{ab} F_{ab}^- \varepsilon_{ij} \epsilon^j + (2w-4) A \eta_i \\
		\delta  B_{ij} &= \overline\epsilon_{(i} \D \Psi_{j)} - \overline\epsilon^k \Lambda_{(i} \varepsilon_{j)k} + 2(1-w) \overline\eta_{(i} \Psi_{j)} \\
		\delta F_{ab}^- &= \frac{1}{4} \varepsilon^{ij} \overline\epsilon_i \D \gamma_{ab} \Psi_j + \frac{1}{4} \overline\epsilon^i \gamma_{ab} \Lambda_i 
			- \frac{1}{2}(1+w) \varepsilon^{ij} \overline\eta_i \gamma_{ab} \Psi_j \\
		\delta \Lambda_i &= -\frac{1}{4} \gamma^{ab} \D (F_{ab}^- \epsilon_i) - \frac{1}{2} \D B_{ij} \varepsilon^{jk} \epsilon_k + \frac{1}{2} C \varepsilon_{ij} \epsilon^j -(1+w) B_{ij} \varepsilon^{jk} \eta_k + \frac{1}{2} (3-w) \gamma^{ab} F_{ab}^- \eta_i \\ 
				\delta C &= -\nabla_m (\varepsilon^{ij} \overline\epsilon_i \gamma^m \Lambda_j) + (2w-4) \varepsilon^{ij} \overline\eta_i \Lambda_j \,.
	\end{align}\label{suconvar}\end{subequations}
	where $\delta$ is a generic $\cN=2$ superconformal transformation being generated by the chiral conformal Killing spinors $\epsilon^i, \eta_i$ and $\epsilon_i, \eta^i$, and we use the matrices $\tau_p^{ij} \equiv \{i\sigma_3, -\mathds{1}_{2 \times 2}, -i\sigma_1\} =: \tau_{pij}^*$. $\gamma_m$ are curved space gamma matrices defined in terms of the vierbein $e_m^a$ and the flat space gamma matrices $\Gamma_a$ as $\gamma_m(x) \equiv e_m^a(x) \Gamma_a$. The conformal Killing spinors satisfy the equations:
	\beq
		\nabla_m \epsilon^i = \gamma_m \eta^i\,, \qquad \nabla_m \epsilon_i = \gamma_m \eta_i \,. \label{CKSE}
	\eeq
	The $osp(2|4)$ transformations can be generated by imposing the following constraints on the conformal Killing spinors:
	\beq
		\eta^j = \frac{i}{2r} \tau_1^{jk} \epsilon_k\,, \qquad \eta_j = \frac{i}{2r} \tau_{1jk} \epsilon^k \,. \label{KSc}
	\eeq
	In the background where the fermions are all vanishing the variations of the bosonic fields automatically vanish. So for the background fields in $\underline\tau_{\cal U}$ we demand that the fermionic variations in (\ref{suconvar}) vanish:
	\begin{align}
		\delta \psi_i =&\; \frac{1}{2} b_{ij} \epsilon^j + \frac{i}{r}(2-w) \tau_\cO \tau_{1ij} \epsilon^j = 0  \\
		\delta \lambda_i =&\; \frac{1}{2} c \epsilon^j \varepsilon_{ij} - \frac{i}{2r}(3-w) b_{ij} \varepsilon^{jk} \tau_{1kl} \epsilon^l = 0\,.
	\end{align}
	These equations are satisfied when:
	\begin{subequations}\begin{align}
		b_{jk} =&\; \frac{2i}{r} (w-2) \tau_{1jk} \tau_{\cal U}~, \\
		c =&\; \frac{2}{r^2} (w-2)(w-3) \tau_{\cal U}~.
	\end{align}\label{coupbck}\end{subequations}
	Now, the product of two chiral multiplets is another chiral multiplet whose bottom component is the product of the bottom components of the individual chiral multiplets and this multiplication is defined in a way such that the integration over the chiral superspace in (\ref{osp24inv}) will simply pick out the top component of the product chiral multiplet $\underline\tau_{\cal U}\, \underline{\cal U}$. The general expression for the top component of $\underline\tau_{\cal U} \, \underline{\cal U}$ is given by: 
	\beq
		 \tau_\cO C + \cO c - \frac{1}{2} \varepsilon^{ik} \varepsilon^{jl} B_{ij} b_{kl} + F_{ab}^- f_{ab}^- + \varepsilon^{ij}\left(\overline \Psi_i \lambda_j + \overline \psi_i \Lambda_j\right).
	\eeq
	When we use the background values where $\tau_{\cal U}$ is a constant, $f_{ab}$ and all the fermions in $\underline\tau_{\cal U}$ vanish and the rest of the fields satisfy (\ref{coupbck}), this becomes:
	\beq
		\tau_{\cal U}\, \cC(x) \equiv \tau_{\cal U} \left[ C(x) + \frac{2}{r^2} (w-2)(w-3) A(x) - \frac{i}{r} (w-2) \tau_1^{ij} B_{ij} (x) \right], \label{topC}
	\eeq
	and (\ref{osp24inv}) reduces to:
	\beq
		\tau_{\cal U} \int_{S^4} \mathrm{d}^4x \sqrt{g}\, \cC(x). \label{osp24deformation}
	\eeq

\section{Ward Identity}
\label{ap:ward}

	For a chiral multiplet $(A, \Psi_i, B_{ij}, F_{ab}^-,\Lambda_i,C)$ of weight $w$ recall the combination~(\ref{topC}):
	\beq
		\cC(x) \equiv C(x) + \frac{2}{r^2} (w-2) (w-3) A(x) - \frac{i}{r} (w-2) \tau_{1}^{ij} B_{ij}(x) \,.
	\eeq
	In this appendix we prove the following identity: If $\mathcal U$ is some $osp(2|4)$ supersymmetric operator, i.e. $\delta_\mathrm{SUSY} {\mathcal U} = 0$, then
	\beq
		\left\langle \left(\int_{S^4} \mathrm{d}^4 x \sqrt{g}\, \cC(x)\right) \mathcal{U} \right\rangle = 32\pi^2 r^2 \langle A(N)\, \mathcal{U} \rangle \,, \label{WardIdentity}
	\eeq
	where $N$ is the North Pole of the sphere. Similarly, for an anti-chiral multiplet we have:
	\beq
		\left\langle{ \mathcal U} \left(\int_{S^4} \mathrm{d}^4 x \sqrt{g}\, \overline\cC(x)\right) \right\rangle = 32\pi^2 r^2 \langle \mathcal{U}\, \overline A(S) \rangle \,, \label{WardIdentityAC}
	\eeq
	where $S$ is the South Pole. 
	\smallskip
	
	From (\ref{CKSE}) and (\ref{KSc}) we see that the nonchiral Killing spinors generating the $osp(2|4)$ algebra preserved on $S^4$  satisfy the equation:
	\beq
		\nabla_m \chi^j = \frac{i}{2r} \gamma_m \chi^j \,, \label{kse}
	\eeq
	where, $\chi^j \equiv \epsilon^j + \tau_1^{jk} \epsilon_k$. In steregraphic coordinates the solutions to (\ref{kse}) are given by:
	\beq
		\chi^j = \frac{1}{\sqrt{1+\frac{x^2}{4r^2}}} \left(\mathds 1 + \frac{i}{2r} x_m \Gamma^m\right) \chi^j_0 \,.
	\eeq
	The constant spinors $\chi_0^j$ parametrize the eight supercharges of $osp(2|4)$. We choose an $su(1|1) \subset osp(2|4)$ by imposing the following constraints:
	\beq
		P_L \chi_0^i = 0, \qquad \chi_0^i = \tau_1^{ij} \varepsilon_{jk} \Gamma_1 \Gamma_2 \chi_0^k \,.
	\eeq
	The chosen Killing spinors and the supersymmetry transformation they generate will henceforth be denoted by $\chi^i$ and $\delta$ respectively. $\chi^i$ satisfy the following equations:
	\begin{align}
		\frac{{\chi_L^i}^\dagger}{\lVert \chi_L \rVert^2} \D(A \chi_R^j) \tau_{2ij} = \nabla_m (U^m A) - \frac{8ir}{x^2}A\,,  \qquad \nabla_m U^m = \frac{8ir}{x^2} - \frac{4i}{r} \,, \label{cond}
	\end{align}
	where we have defined:
	\beq
		\lVert \chi_L \rVert^2 \equiv \lVert \chi_L^1 \rVert^2 = \lVert \chi_L^2 \rVert^2 \quad \mathrm{and,} \quad U^m \equiv \frac{{\chi_L^i}^\dagger \gamma^m \chi_R^j}{\lVert \chi_L \rVert^2} \tau_{2ij} \,.
	\eeq
	\smallskip
	
	Now, using the supersymmetry transformation of a chiral multiplet $(A, \Psi_i, B_{ij}, F_{ab}^-,\Lambda_i,C)$ of weight $w$ (\ref{suconvar}, \ref{KSc}) we can write\footnote{Using (\ref{suconvar}) will also result in some terms proportional to $F_{ab}^-$ and $\nabla_m F_{ab}^-$ in (\ref{BC}), but these terms are vanishing, because while $F_{ab}^-$ is selfdual in Euclidean signature, their coefficients will be proportional to ${\chi^i_L}^\dagger \Gamma^{ab} \gamma^{(r)} \chi^j_{L/R} \tau_{3ij}$, where $\gamma^{(r)}$ is a product of $r$ distinct gamma matrices, and these terms are anti-selfdual as they satisfy: ${\chi_L^i}^\dagger \Gamma^{ab} \gamma^{(r)} \chi_{L/R}^j = {\chi_L^i}^\dagger \Gamma_* \Gamma^{ab} \gamma^{(r)} \chi_{L/R}^j = -\frac{1}{2} {\varepsilon^{ab}}_{cd} {\chi_L^i}^\dagger \Gamma^{ab} \gamma^{(r)} \chi_{L/R}^j$, where $\Gamma_*$ is the chirality matrix.}
	\begin{subequations}\begin{align}
		\frac{1}{2} \tau_1^{ij} B_{ij} \deq&\; \nabla_m(U^m A) - \frac{8ir}{x^2} A - \frac{2i}{r} (w-2) A \label{BCB} \\
		C \deq& - \frac{1}{4} U^m \nabla_m B_{ij} \tau_1^{ij} + \frac{3i}{2r} \tau_1^{ij} B_{ij} + \frac{i}{2r} (w-2) \tau_1^{ij} B_{ij}\,. \label{BCC}
	\end{align}\label{BC}\end{subequations}
	For a chiral multiplet with $w=2$ this calculation was done in more detail in \cite{Gomis:2014woa} and it was shown that we have the following schematic form:
	\beq
		C_{w=2}(x) \deq f(A_{w=2}(x))
	\eeq
	where $f$ is a function that satisfies:
	\beq
		\int_{S^4} \mathrm{d}^4 x \sqrt{g}\, f(A(x)) = 32\pi^2 r^2 A(N) \,.
	\eeq
	We want to repeat this computation now for arbitrary $w$. We define:
	\beq
		\Delta C(x) \equiv C(x) - f(A(x))\,.
	\eeq
	To compute $\Delta C$ we can use (\ref{BCB}) in (\ref{BCC}) to write $C$ entirely in terms of $A$ and then if we consider the expression for $C$ as a polynomial in $(w-2)$ then $\Delta C$ is given by the terms that depend on $(w-2)$. After some simplifications using (\ref{cond}) we find:
	\beq
		\Delta C \deq \frac{2i}{r} (w-2) \nabla_m (U^m A) + \frac{16}{x^2} (w-2) A + \frac{2}{r^2} (w-2)(w-1) A\,.
	\eeq
	Multiplying (\ref{BCB}) by $-\frac{2i}{r}(w-2)$ and adding it to the above equation we find the desired result 
	\beq
		\Delta C + \frac{2}{r^2} (w-2) (w-3) A - \frac{i}{r} (w-2) \tau_1^{ij} B_{ij} \deq 0\,,
	\eeq
	or equivalently:
	\beq
		C(x) + \frac{2}{r^2} (w-2) (w-3) A(x) - \frac{i}{r} (w-2) \tau_1^{ij} B_{ij}(x) = \cC(x) \deq f(A(x)) \,.
	\eeq
	Integrating the two sides of $\deq$ on $S^4$ and putting them inside a correlator with $\mathcal{U}$ gives us the desired identity (\ref{WardIdentity}). The proof of (\ref{WardIdentityAC}) follows similarly.

\section{$tt^*$ Equations from   Sphere Partition Function}
\label{couptt}
In this appendix we prove that the two-point functions in  $SU(N)$ $\mathcal{N}=2$ SQCD (with $2N$ fundamental hypermultiplets) satisfy the coupled $tt^*$ equation. We denote by $\tau,\bar{\tau}$ the marginal coupling which parametrizes the conformal manifold. The chiral ring is generated by the $N-1$ generators \eq{\phi_{k}\propto\Tr(\varphi^k)\ ,\ k=2,...,N}
and a convenient basis for the chiral primaries is
\eq{\cO_i\equiv\cO_{i_2,i_3,...,i_{N}}=\prod_{k=2}^{N}\phi_k^{i_k}\ .}
We will define the matrix of two-point functions on the sphere (dropping the $S^4$ subscript)
 \eq{M_{ab}=\vev{\cO_a(N)\overline{\cO}_b(S)}\ .}
As a consequence of the mixing explained in section~\ref{sec:Toda}, $M_{ab}$ is in general not zero even when $\cO_a$ and $\cO_b$ are not of the same dimension.  The physical  operators $\{\cO'_a\}$ can be obtained by doing a Gram-Schmidt procedure with respect to all the lower-dimensional CPOs (chiral primary operators):
\eq{\cO'_a=\cO_a-\sum_{\Delta_i<\Delta_a}\frac{\vev{\cO_a(N)\overline{\cO'}_i(S)}}{\vev{\cO'_i(N)\overline{\cO'}_i(S)}}\cO'_i\ .}
The physical two-point functions which correspond to the flat space two-point functions are obtained from\eq{G_{ab}=\vev{\cO'_a(N)\overline{\cO'}_b(S)} \,,} which is non zero only if $\Delta_a=\Delta_b$.

\smallskip

We will define the matrix $M^{ij}_{\Delta'}$ to be the inverse of the submatrix of $M_{ij}$ that includes all the operators up to dimension $\Delta'$.
Another useful notation is to denote operators of the form $\phi_2\cO_a$ by $\cO_{\partial a}$, and the corresponding matrix elements are
\eq{M_{\partial i,j}=\vev{\phi_2\cO_i(N)\bar{\cO}_j(S)}\ ,\ M_{i,\partial j}=\vev{\cO_i(N)\overline{\phi_2\cO}_j(S)}\ .}

\smallskip

Derivatives with respect to $\tau,\bar{\tau}$ bring down insertions of $\phi_2,\bar{\phi}_2$ such that the following relations between the matrix elements hold
\eq{&\dtau M_{IJ}=M_{\partial I,J}-M_{10}M_{IJ}\\
&\dbar M_{IJ}=M_{I,\partial J}-M_{01}M_{IJ}}
where \eq{M_{10}=\vev{\phi_2(N)}\ ,\ M_{01}=\vev{\overline{\phi_2}(S)}\ .}
In the proceeding of this section, 
we will use the indices $a,b,c$ to denote operators of dimension $\Delta$, indices $i,j,k,l$ to denote operators of dimension smaller than $\Delta$ and $I,J,K$ to denote operators up to dimension $\Delta$. Contracted indices are summed over all their possible values unless specified differently.
Due to the Gram-Schmidt procedure, we can  write $G_{ab}$ in the following way ($\Delta_a=\Delta_b=\Delta$)
\eq{G_{ab}=M_{ab}-M_{ai}M^{ij}_{\Delta-2}M_{jb}\ .} 
It will be useful to show that the inverse of $G_{ab}$ denoted by $G^{bc}$ is equal to $G^{bc}=M_{\Delta}^{bc}$.
Proof:
\eq{&G_{ab}G^{bc}=(M_{ab}-M_{ai}M_{\Delta-2}^{ij}M_{jb})M_{\Delta}^{bc}=M_{ab}M_{\Delta}^{bc}-M_{ai}M_{\Delta-2}^{ij}M_{jb}M_{\Delta}^{bc}\\
&=M_{ab}M_{\Delta}^{bc}+M_{ai}M_{\Delta-2}^{ij}M_{jk}M_{\Delta}^{kc}=M_{ab}M_{\Delta}^{bc}+M_{ai}M_{\Delta}^{ic}=\delta_a^c\ .}
The $tt^*$ equations \rf{tt*} in the holomorphic gauge and in these notations take the form
\eql{ttstar}{\dbar(\dtau G_{ab}G^{bc})=G_{\partial a,\partial b}G^{bc}-G_2\delta_a^c-G_{a\partial i}G^{i j}\delta_{\partial j}^c\ .}
In order to prove \eqref{ttstar}, we need to compute $\dbar(\dtau G_{ab}G^{bc})$. Do it in steps:

\eq{&\dtau G_{ab}=\dtau(M_{ab}-M_{ai}M_{\Delta-2}^{ij}M_{jb})\\&=M_{\partial a,b}-M_{10}M_{ab}-M_{\partial a,i}M_{\Delta-2}^{ij}M_{jb}-M_{ai}M_{\Delta-2}^{ij}M_{\partial j,b}+\\&+M_{ai}M_{\Delta-2}^{ik}M_{\partial k,l}M_{\Delta-2}^{lj}M_{jb}+M_{10}M_{ai}M_{\Delta-2}^{ij}M_{jb}M_{ai}M_{\Delta-2}^{ij}M_{jb}\\
&=M_{\partial a,b}-M_{10}G_{ab}-M_{\partial a,i}M_{\Delta-2}^{ij}M_{jb}-M_{ai}M_{\Delta-2}^{ij}M_{\partial j,b}+M_{ai}M_{\Delta-2}^{ik}M_{\partial k,l}M_{\Delta-2}^{lj}M_{jb}\\
&=M_{\partial a,b}-M_{10}G_{ab}-M_{\partial a,i}M_{\Delta-2}^{ij}M_{jb}-\sum_{\partial k\in \Delta}M_{ai}M_{\Delta-2}^{ik}M_{\partial k,b}+\sum_{\partial k\in \Delta}M_{ai}M_{\Delta-2}^{ik}M_{\partial k,l}M_{\Delta-2}^{lj}M_{jb}\\
&=M_{\partial a,b}-M_{10}G_{ab}-M_{\partial a,i}M_{\Delta-2}^{ij}M_{jb}-\sum_{\partial k\in \Delta}M_{ai}M_{\Delta-2}^{ik}G_{\partial k,b}}
and
\eq{&\dtau G_{ab}G^{bc}=\left(M_{\partial a,b}-M_{10}G_{ab}-M_{\partial a,i}M_{\Delta-2}^{ij}M_{jb}-\sum_{\partial j\in \Delta}M_{ai}M_{\Delta-2}^{ij}G_{\partial j,b}\right)M_{\Delta}^{bc}\\
&=M_{\partial a,b}M_{\Delta}^{bc}-M_{10}\delta_a^c+M_{\partial a,i}M_{\Delta-2}^{ij}M_{jk}M_{\Delta}^{kc}-\sum_{\partial j\in \Delta}M_{ai}M_{\Delta-2}^{ij}\delta_{\partial j}^c\\
&=M_{\partial a,b}M_{\Delta}^{bc}-M_{10}\delta_a^c+M_{\partial a,i}M_{\Delta}^{ic}-\sum_{\partial j\in \Delta}M_{ai}M_{\Delta-2}^{ij}\delta_{\partial j}^c\ .}
and finally
\eq{&\dbar(\dtau G_{ab}G^{bc})=\dbar\left(M_{\partial a,b}M_{\Delta}^{bc}-M_{10}\delta_a^c+M_{\partial a,i}M_{\Delta}^{ic}-\sum_{\partial j\in \Delta}M_{ai}M_{\Delta-2}^{ij}\delta_{\partial j}^c\right)\ .}
Compute the different terms:
\eq{&\dbar M_{\partial a,I}M_{\Delta}^{Ic}=M_{\partial a,\partial I}M_{\Delta}^{Ic}-M_{01}M_{\partial a,I}M_{\Delta}^{Ic}-M_{\partial a,I}M_{\Delta}^{IJ}(M_{J,\partial K}-M_{01}M_{JK})M^{Kc}\\
&=M_{\partial a,\partial I}M_{\Delta}^{Ic}-M_{01}M_{\partial a,I}M_{\Delta}^{Ic}-M_{\partial a,I}M_{\Delta}^{IJ}M_{J,\partial K}M^{Kc}+M_{\partial a,I}M_{\Delta}^{IJ}M_{01}M_{JK}M^{Kc}\\
&=M_{\partial a,\partial b}M_{\Delta}^{bc}+M_{\partial a,\partial i}M_{\Delta}^{ic}-M_{\partial a,\partial i}M_{\Delta}^{ic}-M_{\partial a,I}M_{\Delta}^{IJ}M_{J\partial b}M_{\Delta}^{bc}=G_{\partial a,\partial b}G^{bc}\ .}
Second term:
\eq{\dbar(-M_{10}\delta_a^c)=-(M_{11}-M_{10}M_{01})\delta_a^c=-G_2\delta_a^c\ .}
Last term:
\eq{&-\sum_{\partial j\in\Delta}\delta_{\partial j}^c\dbar(M_{ai}M_{\Delta-2}^{ij})=-\sum_{\partial j\in\Delta}\delta_{\partial j}^c(M_{a\partial i}M_{\Delta-2}^{ij}-M_{al}M_{\Delta-2}^{lk}M_{k,\partial i}M_{\Delta-2}^{ij})\\
&=-\sum_{\partial i,\partial j\in\Delta}\delta_{\partial j}^c(M_{a\partial i}M_{\Delta-2}^{ij}-M_{al}M_{\Delta-2}^{lk}M_{k,\partial i}M_{\Delta-2}^{ij})=-\sum_{\partial j\in\Delta}\delta_{\partial j}^cG_{a\partial i}G^{ij}\ .}
Putting everything together we get exactly \eqref{ttstar}.

\section{Scheme Independence of the Results}
The sphere partition function is subject to K\"ahler ambiguity transformations 
\begin{equation}
\ln Z[{S^4}]\to\ln Z[{S^4}]+f(\tau^i)+\bar f(\bar{\tau}^{\bar i})\;.
\end{equation}
That is, sphere partition functions that were computed in different regularization schemes may differ by holomorphic functions in the exactly marginal couplings \cite{Gomis:2014woa}. 
More generally, the deformed partition function $Z[{S^4}](\tau^i,\bar\tau^{\bar i},\tau^A,\bar\tau^{\bar A})$ is subject to holomorphic ambiguities, as discussed in section \ref{relation}.

The expressions obtained for the extremal correlators in our prescription are, by construction, unambiguous. The effect of the holomorphic ambiguities on sphere correllators is in holomorphic contributions to the mixing of chiral primaries with lower dimensional chiral primaries (see equations (\ref{mixingSUSY}-\ref{16})), and the Gram-Schmidt procedure subtracts these holomorphic contributions.  The algorithm described in section \ref{summary} is therefore guaranteed to yield results that are scheme independent. 
Here we would like to demonstrate how this works.

Let us start with the example of gauge group $SU(2)$. Using the recursive formula (\ref{rec}), the invariance of the extremal two-point functions follows from the invariance of the boundary condition $G_2=16\,\partial_\tau \partial_{\bar\tau} \ln Z[S^4]$ under K\"ahler transformations. Alternatively, consider the formula (\ref{generaldet}) and note that 
\begin{equation}
\partial_\tau^l\partial_{\bar{\tau}}^j\left(e^{f(\tau)}Z[S^4]\right)=e^{f(\tau)}\partial_\tau^l\partial_{\bar{\tau}}^jZ[S^4]+\sum_{k=0}^{l-1}\mat{l\\k}\left(\partial_\tau^{l-k}e^{f(\tau)}\right)\partial_\tau^k\partial_{\bar{\tau}}^jZ[S^4]\;.
\end{equation}
The second term in the right hand side of the equation above is a linear combination of the first $l$ columns of the matrix defined by the first term, and therefore does not affect the determinant,
\begin{equation}
\det\Big(\partial_\tau^l\partial_{\bar{\tau}}^j\left(e^{f(\tau)}Z[S^4]\right)\Big)=\det\Big( e^{f(\tau)}\partial_\tau^l\partial_{\bar{\tau}}^jZ[S^4]\Big)\;.
\end{equation}
It follows that equation (\ref{generaldet}) is invariant under holomorphic transformations (and similarly under antiholomorphic transformations.)

More generally, every extremal two-point function that we would like to compute is given in our prescription in terms of determinants of the Gram-Schmidt matrix of two-point functions on the sphere. The holomorphic mixing can always be canceled by subtracting from columns linear combinations of the previous columns, and therefore the holomorphic ambiguities do not affect the (appropriately normalized) determinants.
 Importantly, non-holomorphic contributions to $\ln Z[S^4]$, such as the one due to the anomaly discussed in \cite{Gomis:2015yaa}, do not simply mix columns and rows with the previous ones, and they do affect the result of the Gram-Schmidt procedure.

\end{fmffile}
\bibliography{refs}
\end{document}